\PassOptionsToPackage{x11names}{xcolor}
\documentclass[twocolumn, dvipsnames, trackchanges]{aastex702}
\usepackage[autostyle, english = american]{csquotes}
\MakeOuterQuote{"}
\usepackage{hyperref}
\usepackage{xcolor}
\hypersetup{
    colorlinks=true,
    linkcolor=DeepSkyBlue3,
    filecolor=DeepSkyBlue3,      
    urlcolor=DeepSkyBlue3,
    citecolor=DeepSkyBlue3,
    }
\usepackage[T1]{fontenc}
\usepackage{graphicx}
\usepackage{amsmath}
\usepackage{listings}
\newcommand{\flc}{\texttt{flc}}

\begin{document}

\title{The JWST Resolved Stellar Populations Early Release Science Program. X. Cross-Observatory Proper Motion Measurements with HST and JWST}

\author[0000-0003-1634-4644]{Jack T. Warfield}
\affiliation{Department of Astronomy, University of California, Berkeley, CA 94720, USA}
\affiliation{Department of Astronomy, University of Virginia, 530 McCormick Road, Charlottesville, VA 22904, USA}
\email[show]{warfield@berkeley.edu}

\author[0000-0002-3204-1742]{Nitya Kallivayalil}
\affiliation{Department of Astronomy, University of Virginia, 530 McCormick Road, Charlottesville, VA 22904, USA}
\email{njk3r@virginia.edu}

\author[0000-0001-8368-0221, sname='Sohn']{Sangmo Tony Sohn}
\affiliation{Space Telescope Science Institute, 3700 San Martin Drive, Baltimore, MD 21218, USA}
\email{tsohn@stsci.edu}

\author[0000-0001-9061-1697]{Christopher T. Garling}
\affiliation{Space Telescope Science Institute, 3700 San Martin Drive, Baltimore, MD 21218, USA}
\email{cgarling@stsci.edu}

\author[orcid=0000-0002-6021-8760]{Andrew B. Pace}
\thanks{Galaxy Evolution and Cosmology (GECO) Fellow}
\affiliation{Department of Astronomy, University of Virginia, 530 McCormick Road, Charlottesville, VA 22904, USA}
\email{yyr6cm@virginia.edu}

\author[0000-0001-7827-7825]{Roeland P. van der Marel}
\affiliation{Space Telescope Science Institute, 3700 San Martin Drive, Baltimore, MD 21218, USA}
\affiliation{Center for Astrophysical Sciences, The William H. Miller III Department of Physics \& Astronomy, Johns Hopkins University, Baltimore, MD 21218, USA}
\email{marel@stsci.edu}

\author[0000-0002-6442-6030]{Daniel~R.~Weisz}\affil{Department of Astronomy, University of California, Berkeley, CA 94720, USA}\affil{Miller Institute for Basic Research, University of California Berkeley, Berkeley, CA, 94720, USA}\email{dan.weisz@berkeley.edu}

\author[0000-0003-4207-3788]
{Mark A.\ Fardal}
\affiliation{Eureka Scientific, 2452 Delmer St., Suite 100, Oakland, CA 94602, USA}
\email{mfardal@eurekasci.com}

\author[0009-0002-1233-2013]{Niusha Ahvazi}
\thanks{Galaxy Evolution and Cosmology (GECO) Fellow}
\affiliation{Department of Astronomy, University of Virginia, 530 McCormick Road, Charlottesville, VA 22904, USA}
\email{nahvazi@virginia.edu}

\author[0000-0003-2861-3995]{Jay Anderson}
\affiliation{Space Telescope Science Institute, 3700 San Martin Drive, Baltimore, MD 21218, USA}
\email{jay@stsci.edu}

\author[0000-0001-9649-8103]{Kaia R. Atzberger}
\affiliation{Department of Astronomy, University of Virginia, 530 McCormick Road, Charlottesville, VA 22904, USA}
\email{yzb4en@virginia.edu}

\author[0000-0001-8354-7279]{Paul Bennet}
\affiliation{Space Telescope Science Institute, 3700 San Martin Drive, Baltimore, MD 21218, USA}
\email{pbennet@stsci.edu}

\author[0000-0003-4850-9589]{Martha L. Boyer}
\affiliation{Space Telescope Science Institute, 3700 San Martin Drive, Baltimore, MD 21218, USA}
\email{mboyer@stsci.edu}

\author[0000-0002-2970-7435]{Roger E. Cohen}
\affiliation{Eureka Scientific, 2452 Delmer St., Suite 100, Oakland, CA 94602, USA}
\affiliation{Department of Physics and Astronomy, Rutgers, The State University of New Jersey, 136 Frelinghuysen Rd, Piscataway, NJ 08854, USA}
\email{cohenrogere@gmail.com}

\author[0000-0003-0303-3855]{Andrew A. Cole}
\affiliation{School of Natural Sciences, University of Tasmania, P.O. Box 807, Sandy Bay, TAS 7006 Australia}
\email{andrew.cole@utas.edu.au}

\author[0000-0001-6464-3257]{Matteo Correnti}
\affiliation{INAF Osservatorio Astronomico di Roma, Via Frascati 33, 00078, Monteporzio Catone, Rome, Italy}
\affiliation{ASI-Space Science Data Center, Via del Politecnico, I-00133, Rome, Italy}
\email{matteo.correnti@inaf.it}

\author[0000-0001-8416-4093]{Andrew E. Dolphin}
\affiliation{Raytheon Technologies, 1151 E. Hermans Road, Tucson, AZ 85756, USA}
\affiliation{Steward Observatory, University of Arizona, 933 N. Cherry Avenue, Tucson, AZ 85719, USA}
\email{adolphin@raytheon.com}

\author[0000-0002-7007-9725]{Marla C. Geha}
\affiliation{Department of Astronomy, Yale University, New Haven, CT 06520, USA}
\email{marla.geha@yale.edu}

\author[0000-0002-5581-2896]{Mario Gennaro}
\affiliation{Space Telescope Science Institute, 3700 San Martin Drive, Baltimore, MD 21218, USA}
\affiliation{Center for Astrophysical Sciences, The William H. Miller III Department of Physics \& Astronomy, Johns Hopkins University, Baltimore, MD 21218, USA}
\email{gennaro@stsci.edu}

\author[0000-0001-7494-5910]{Kevin A. McKinnon}
\affiliation{David A. Dunlap Department of Astronomy \& Astrophysics, University of Toronto, 50 St George Street, Toronto ON M5S 3H4, Canada}
\email{kevin.mckinnon@utoronto.ca}

\author[0000-0001-5538-2614]{Kristen B. W. McQuinn}
\email{kmcquinn@stsci.edu}
\affiliation{Space Telescope Science Institute, 3700 San Martin Drive, Baltimore, MD 21218, USA}
\affiliation{Department of Physics and Astronomy, Rutgers, The State University of New Jersey, 136 Frelinghuysen Rd, Piscataway, NJ 08854, USA}
\affiliation{Center for Astrophysical Sciences, The William H. Miller III Department of Physics \& Astronomy, Johns Hopkins University, Baltimore, MD 21218, USA}

\author[0000-0002-8092-2077]{Max J. B. Newman}
\affiliation{Space Telescope Science Institute, 3700 San Martin Drive, Baltimore, MD 21218, USA}
\email{mnewman@stsci.edu}

\author[0000-0002-9820-1219]{Ekta~Patel}
\affiliation{Department of Astrophysics and Planetary Sciences, Villanova University,  800 E. Lancaster Ave, Villanova, PA 19085, USA}
\email{ekta.patel@villanova.edu}

\author[0000-0002-4378-8534]{Karin M. Sandstrom}
\affiliation{Center for Astrophysics and Space Sciences, Department of Physics, University of California San Diego, 9500 Gilman Drive, La Jolla, CA 92093, USA}
\email{kmsandstrom@ucsd.edu}

\author[0000-0002-1445-4877]{Alessandro Savino}
\affiliation{Department of Astronomy, University of California, Berkeley, CA 94720, USA}
\email{astrosavino@gmail.com}

\author[0000-0003-0605-8732]{Evan D. Skillman}
\affiliation{Minnesota Institute for Astrophysics, University of Minnesota,
116 Church Street SE, Minneapolis, MN 55455, USA}
\email{skill001@umn.edu}

\author[0000-0002-2732-9717]{Eduardo Vitral}
\affiliation{LIRA, Observatoire de Paris, Universit\'e PSL, Sorbonne Universit\'e, Universit\'e Paris Cit\'e, CY Cergy Paris Universit\'e, CNRS, 92195 Meudon, France}
\email{eduardo.vitral@obspm.fr}

\author[0000-0002-1343-134X]{Laura L. Watkins}
\affiliation{AURA for the European Space Agency (ESA), Space Telescope Science Institute, 3700 San Martin Drive, Baltimore, MD 21218, USA}
\email{lwatkins@stsci.edu}

\author[0000-0002-7502-0597]{Benjamin F. Williams}
\affiliation{Department of Astronomy, University of Washington, Box 351580, U.W., Seattle, WA 98195-1580, USA}
\email{ben@astro.washington.edu}

\collaboration{all}{The JWST Resolved Stellar Populations ERS and HSTPROMO Collaborations}

\begin{abstract}

We present the first absolute systemic proper motion (PM) of a galaxy measured by combining Hubble Space Telescope (HST) and JWST imaging. The PM of WLM, an isolated dwarf galaxy at $\sim$1~Mpc, is measured against background galaxies with Advanced Camera for Surveys (ACS) F814W imaging from 2015 and 2024 and NIRCam imaging from 2022 and 2024. The two ACS epochs give a systemic PM of $(\mu_{\alpha}^*, \mu_{\delta}) = (+0.087 \pm 0.019, -0.079 \pm 0.017)$~mas~yr$^{-1}$ from 145 galaxies over 8.93~years, which agrees with previous Gaia-based PMs to within 0.4$\sigma$ and is the most precise PM of WLM. For the HST-to-JWST PMs, we fit each NIRCam detector separately, match the point-spread functions with empirically derived convolution kernels, and correct biases with injected galaxies. With an F090W second epoch, the PM is $(+0.081 \pm 0.045, -0.113 \pm 0.046)$~mas~yr$^{-1}$ from 43 galaxies over 7.02~years, which, like the F150W PM, agrees with HST-to-HST over the same area to within 1$\sigma$, while F200W differs by $\sim$2$\sigma$. The long-wavelength F250M and F430M PMs are consistent to within 1$\sigma$, though less precise. Fewer galaxies can be used at wavelengths farther from F814W. A cross-observatory systematic of $\sim$0.05~mas~yr$^{-1}$ cannot be ruled out over 7~years. At equal baselines, an F090W second epoch is at least as precise as our second ACS epoch over the same area. We recommend the NIRCam filter closest in wavelength to the first-epoch filter, though the long-wavelength filters show promise for closer targets or longer baselines.

\end{abstract}


\section{Introduction}

Galaxy dynamics, and their role in characterizing dark matter, are directly important for addressing the tensions between dark matter theory and observations \citep[e.g.,][]{Bullock2017}.
As one example, velocity measurements can be used to estimate the mass of a system's dark matter halo \citep[e.g.,][]{Wolf2010, Patel2017b}.
Accurate halo mass estimates are important for comparing observed galaxies to those produced by models and for testing the phenomenological predictions of models. However, line-of-sight velocities, measured from Doppler shifts, provide only one component of a galaxy's motion. The other two come from its absolute proper motion (PM), the angular velocity of the galaxy as a whole on the plane of the sky, measured against an inertial reference frame. This is distinct from the relative PMs of the stars within a galaxy with respect to each other, which describe its internal kinematics but not its orbit. With all three components, the orbits of satellite galaxies can be reconstructed, which in turn constrain the masses of their hosts (e.g., \citealt{Patel2017b, Patel2023}), the timing of their infall relative to when they stopped forming stars (e.g., \citealt{Battaglia2022, Bennet2024, Bennet2025}), and whether satellites that appear to lie in planes are in fact co-orbiting (e.g., \citealt{Sohn2020, CasettiDinescu2024}; see also \citealt{Weisz2026}).

Andromeda (M31), its satellites, and the galaxies of the Local Volume and beyond are the next frontier for probing these questions \citep[e.g.,][]{Patel2023, Savino2025}. While about 90\% of the known Milky Way (MW) satellites have measured PMs, mostly from Gaia (e.g., \citealt{McConnachie2020b, Battaglia2022, Pace2022}), M33 is the only M31 satellite with enough bright stars for Gaia to measure its systemic PM \citep{vanderMarel2019, Battaglia2022}. As of 2026 November, only nine of the approximately forty known M31 satellites have published absolute PMs (\citealt{CasettiDinescu2025, Bennet2025}; see also the compilation of \citealt{Pace2025}). Those of M33 and IC~10 were first measured with very long baseline interferometry of water masers \citep{Brunthaler2005, Brunthaler2007}. M33 has since also been measured with Gaia \citep{vanderMarel2019}, and IC~10 with the Hubble Space Telescope (HST) and Gaia combined \citep{Bennet2024}. The other seven rest on multi-epoch HST imaging: NGC~147 and NGC~185 \citep{Sohn2020}, Andromeda~VII \citep{Warfield2023}, Andromeda~III \citep{CasettiDinescu2024}, Andromeda~V and VI \citep{CasettiDinescu2025}, and Pisces~I (LGS~3; \citealt{Bennet2025}). For about twenty of the remaining satellites, the first deep Advanced Camera for Surveys (ACS) Wide Field Channel (WFC) imaging suited to PM measurements was acquired between 2019 and 2021 (HST-GO 15902, PI D. Weisz; \citealt{Savino2022, Savino2025}). \citet{Sohn2020} reached PM uncertainties of $\sim$0.014~mas~yr$^{-1}$ per coordinate ($\sim$45--50~km~s$^{-1}$ at their distances) for NGC~147 and NGC~185, two of the brightest M31 satellites, with $\sim$8-year baselines. The remaining satellites lie at similar distances ($\sim$500--900~kpc; \citealt{Savino2022}), but they are fainter and their first-epoch imaging is shallower, and so we expect them to need baselines of more than 10~years to reach the same precision. Second-epoch imaging may therefore not be possible within HST's remaining mission.

Therefore, cross-observatory measurements are crucial for the future of absolute PMs of dwarf galaxies in the Andromeda system and beyond. JWST's Near Infrared Camera (NIRCam) is the obvious candidate for this second-epoch imaging, as it matches or exceeds the angular resolution and sensitivity of ACS/WFC and is expected to remain in operation into the 2040s. However, its ability to measure absolute PMs in tandem with archival HST imaging has remained largely untested. Between two epochs of HST imaging, PM systematics can be mitigated relatively simply (for instance, by duplicating the original pointing and orientation; e.g., \citealt{Sohn2012}). JWST, however, operates with different filters, detectors, dither patterns, and orientation constraints. The effects of these differences on cross-observatory systematics must be studied in order to realize JWST's full potential for PM experiments. Although \citet{Libralato2023} and \citet{Ziliotto2023} have published PMs combining ACS/WFC data with NIRISS and NIRCam data, respectively (see also \citealt{Ziliotto2026}), these are relative PMs of the stars within a field with respect to each other. They can only be put on an absolute scale (linked to a universal reference frame, against which the motions of other fields can be compared) by referring back to Gaia or to some other inertial reference frame (Section~\ref{sec:pms}).

In this paper, we present a modification of the galaxy spread function template-fitting procedure detailed in \citet{Warfield2026}, and apply it to imaging from HST's ACS/WFC and JWST's NIRCam to measure the absolute PM of the isolated Local Group (LG) galaxy WLM. This is the first absolute systemic PM of a galaxy to be measured by combining HST and JWST imaging. These measurements are meant to serve as a proof of concept for the feasibility of using NIRCam observations as a second epoch to HST observations for PM measurements, and to set the stage for generalizing this process to cross-observatory measurements with other facilities, such as ESA's Euclid \citep{Euclid2025} and the recently launched Nancy Grace Roman Space Telescope \citep{WFIRST2019}.

WLM was imaged with NIRCam in 2022 by the JWST Resolved Stellar Populations Early Release Science (ERS) Program (PID 1334; \citealt{Weisz2023}), and has archival HST/ACS imaging from 2015. An isolated, star-forming dwarf irregular galaxy ($M_V \approx -14.4$; \citealt{Pace2025}) at a distance of $\sim$1~Mpc \citep{Yan2025}, WLM is unlikely to have ever interacted with the MW, and could only have interacted with M31 if M31 is relatively massive \citep{McConnachie2021}. Orbits based on its existing PMs place it on its first passage through the LG \citep{Bennet2024}, and it was likely even more isolated in the past \citep{Battaglia2022, Cohen2025}. At its distance, WLM is a good proxy for the PM precision possible for distant systems such as M31 satellites and those beyond the LG. Its PM has so far only been measured using Gaia, either alone or combined with one epoch of HST imaging \citep{McConnachie2021, Battaglia2022, Bennet2024}, with uncertainties of $\gtrsim0.02$~mas~yr$^{-1}$ ($\gtrsim$90~km~s$^{-1}$ at its distance). We obtained a third epoch of imaging in 2024 with both HST and JWST (HST-GO 17473 and JWST-GO 4570, PI J. Warfield; Section~\ref{sec:data}), so that HST-to-HST and HST-to-JWST PMs can be compared for the same fields and the same background galaxies.

This paper is organized as follows. In Section~\ref{sec:pms}, we review what a PM measurement requires. In Section~\ref{sec:data}, we describe the HST and JWST imaging of WLM and our preliminary source catalogs. In Section~\ref{sec:methods}, we describe our methodology, in the order of the steps needed to reproduce our measurements, along with the alternatives we tested at each step. We present our PM results and compare the second epochs in Section~\ref{sec:results}, discuss what they imply for future cross-observatory measurements in Section~\ref{sec:discussion}, and summarize our conclusions in Section~\ref{sec:conclusions}.

\section{General Requirements for Measuring Proper Motions} \label{sec:pms}

A PM measurement requires at least two sets of positions, or epochs, separated by a temporal baseline, $\Delta t$. The PM is then $\mu=\Delta x/\Delta t$, where $\Delta x$ is the angular change in position between the epochs, measured relative to sources that are either stationary or whose own motions are already known.

For galaxies in and around the LG, these changes in position are very small, which makes PM measurements technically difficult.
Draco~II, the closest confirmed dwarf satellite of the MW\footnote{Ursa~Major~III/UNIONS~1 ($\sim$10~kpc; \citealt{Smith2024}) is closer, but has not yet been confirmed as a dwarf galaxy.} at a distance of about 22~kpc \citep{Longeard2018, Pace2025}, would take about 36~years to move the length of one $50\times50$~mas pixel of HST's ACS/WFC detectors \citep[assuming the PM as reported by][]{Pace2022, Battaglia2022, Warfield2026}.
WLM, at $\sim$1~Mpc, has a PM of $\sim$0.1~mas~yr$^{-1}$ (Section~\ref{ssec:systemic}), and would take more than 400~years.
Therefore, PM measurements require us to (1) have long temporal baselines and/or (2) be able to place both epochs of imaging into the same reference frame with extremely high (sub-pixel) precision.

For two-epoch space-based imaging, this reference frame is set by sources in the field whose own motions are either negligible or already known. Quasi-stellar objects (QSOs) make good reference objects, as their profiles are similar to those of stars (e.g., in the PM study of the LMC \& SMC;  \citealt{Kallivayalil2013}). However, there are few confirmed QSOs in any given field, and often none. Stars with Gaia PMs can also be used, by tying the field to the Gaia reference frame (e.g., \citealt{delPino2022, Warfield2023, McKinnon2024}). This needs no model of each reference source beyond the point-spread function (PSF), but its precision is set by the number of Gaia stars in the field and by the uncertainties of their Gaia PMs \citep{Warfield2023}, and a single ACS/WFC field often contains few of them (e.g., nine and two in the fields of Andromeda~V and VI; \citealt{CasettiDinescu2025}). This approach will improve with future Gaia data releases.

The most consistently available reference sources are background galaxies, which have no measurable PMs of their own and which number in the hundreds in a single deep HST pointing. For most of the M31 satellites, and for galaxies beyond the LG, they are also the only option. At these distances, the tip of the red giant branch (RGB) is at or beyond Gaia's limiting magnitude, and so only galaxies with young or intermediate-age stars above the tip of the RGB have any members with Gaia astrometry \citep{Battaglia2022}. The few foreground stars with Gaia PMs in each field cannot set the reference frame precisely enough at present (e.g., \citealt{CasettiDinescu2025}), and many of them saturate in the long exposures that these targets need \citep{Warfield2023}. The background galaxy technique therefore has to work across observatories if PMs are to be measured for these systems. However, background galaxies have unique, complex shapes that are non-trivial to model. Those with compact enough nuclei can be fit approximately with PSFs (e.g., \citealt{Kalirai2004, Bedin2006}), but a more promising and proven (but also more intensive) approach is template fitting. In this approach, the profile of each background galaxy is built empirically from a super-sampled stack of all images in an epoch, and this template is then used to fit for the galaxy's position in each individual image. \citet{Mahmud2008} were the first to explore this method, and showed, using the Hubble Ultra Deep Field, that a single 1200~s ACS/WFC exposure contains enough measurable galaxies to define a reference frame to better than 0.02~pixels (1~mas). Template fitting has since been used to measure the PMs of M31 \citep{Sohn2012}, Leo~I \citep{Sohn2013}, Draco and Sculptor \citep{Sohn2017}, NGC~147 and NGC~185 \citep{Sohn2020}, Leo~A and the Sagittarius dwarf irregular \citep{Bennet2024}, six other LG dwarfs \citep{Bennet2025}, and Draco~II \citep{Warfield2026}, and background galaxies have been used alongside QSOs for Draco \citep{Pryor2015} and Leo~II \citep{Piatek2016}.

However, this reference frame can also be vulnerable to systematics. Building a template requires very deep observations, both in exposure time per image and in the number of dithered images per epoch, to reach an adequate signal-to-noise ratio (SNR) for sampling each galaxy's profile. Because background galaxies are, at best, only as bright as the faintest stars in a target galaxy, their centroids are also more prone to systematic offsets from charge-transfer (in)efficiency (CTE) in the ACS/WFC detectors \citep{Anderson2018}. These offsets can be corrected for by using stars near a given background galaxy on the detector (e.g., \citealt{Sohn2020, Warfield2026}), but this correction is difficult-to-impossible to apply for many background galaxies in the fields of faint, angularly compact, and/or sparse targets. Template fits are also sensitive to the PSF itself: \citet{Mahmud2008} found that, because the PSF is asymmetric, observations taken at different orientations can shift the measured positions of stars and galaxies at the $\sim$0.03~pixel level. Finally, a template is only an empirical description of how a galaxy appears through one telescope, detector, and filter. Because galaxy morphologies change with wavelength, and because the PSFs of HST and JWST differ, a template built from one observatory's images is not guaranteed to describe the same galaxy as seen by the other. Taken all together, background galaxies are the one set of reference sources available in nearly any deep field, including those of the M31 satellites, but they are also the reference sources whose systematics most need to be understood before they can be used across observatories.

\section{Data} \label{sec:data}

In this section, we describe the HST and JWST imaging of WLM and the preliminary source catalogs that we build from it. Table~\ref{tab:obs} summarizes the observations. All of the HST and JWST data used in this paper can be found in MAST: \dataset[doi:10.17909/akpq-yt43]{https://doi.org/10.17909/akpq-yt43}.

\begin{deluxetable*}{llllccccll}
\tablewidth{0pt}
\tabletypesize{\footnotesize}
\tablecaption{HST and JWST Observations Used in This Work\label{tab:obs}}
\tablehead{
\colhead{Epoch} & \colhead{Instrument} & \colhead{Program (PI)} & \colhead{Date} & \colhead{Filter} & \colhead{$N_{\rm exp}$} & \colhead{$t_{\rm exp}$} & \colhead{$N_{\rm grp} \times N_{\rm int}$} & \colhead{Detectors} & \colhead{$\Delta t$} \\
\colhead{} & \colhead{} & \colhead{} & \colhead{(UT)} & \colhead{} & \colhead{} & \colhead{(s)} & \colhead{} & \colhead{Used} & \colhead{(yr)}
}
\startdata
E1 & ACS/WFC & HST-GO 13768 (Weisz) & 2015 Jul 17--19 & F814W & 26 & 1140--1410 & \nodata & \nodata & \nodata \\
E1 & ACS/WFC & HST-GO 13768 (Weisz) & 2015 Jul 17--19 & F475W\tablenotemark{a} & 24 & 1140 & \nodata & \nodata & \nodata \\
E2 & NIRCam & JWST-ERS 1334 (Weisz) & 2022 Jul 23--24 & F090W/F430M & 4 & 7537 & $8 \times 9$ & B1--B3, B5 & 7.02 \\
E2 & NIRCam & JWST-ERS 1334 (Weisz) & 2022 Jul 24 & F150W/F250M & 4 & 5862 & $8 \times 7$ & B1--B3, B5 & 7.02 \\
E3 & ACS/WFC & HST-GO 17473 (Warfield) & 2024 Jun 21 & F814W & 6 & 1189--1225 & \nodata & \nodata & 8.93 \\
E3 & NIRCam & JWST-GO 4570 (Warfield) & 2024 Jul 22 & F200W/F360M & 4 & 1460 & $7 \times 2$ & A3, A4, B3, B4, A5, B5 & 9.02 \\
\enddata
\tablecomments{$N_{\rm exp}$ is the number of exposures used. For NIRCam, the filter column gives the short-wavelength/long-wavelength (SW/LW) filter pair observed simultaneously, $t_{\rm exp}$ is the effective exposure time per image, and all observations used the \texttt{MEDIUM8} read-out pattern with a four-point sub-pixel dither (\texttt{4-POINT-MEDIUM-WITH-NIRISS} for E2, \texttt{STANDARD} for E3). $\Delta t$ is the time between the start of the first exposure of each data set and that of the first E1 F814W exposure (from their \texttt{EXPSTART} values), which is the baseline used for the PMs.}
\tablenotetext{a}{Used only for selecting likely member stars from the color--magnitude diagram, not for the PM analysis.}
\end{deluxetable*}

\subsection{HST ACS/WFC Imaging}

WLM's first-epoch (E1) ACS observations were taken in 2015 July with HST/ACS's F814W filter,\footnote{Wide-$I$, $\sim$700--950~nm.} resulting in twenty-six \flc\ images\footnote{The \flc\ filetype indicates calibrated, CTE-corrected, and flat-fielded single-exposure images with no geometric distortion correction applied.} with varying exposure times of 1140, 1230, or 1410~s each (HST-GO 13768, PI D. Weisz). These observations were taken along with observations in ACS's F475W filter. These F475W observations are used to select likely WLM members from the color--magnitude diagram (CMD; Section~\ref{ssec:catalogs}), but are not used for the PM analysis. Second-epoch (which we will refer to as ``E3'') ACS observations were taken in 2024 June (HST-GO 17473, PI J. Warfield), providing a nine-year baseline between epochs. These observations were planned as eight exposures over four orbits to execute a full eight-point sub-pixel dither. However, two of these exposures were not taken due to guide-star acquisition failure, and each exposure time was cut short by $\sim$100~s due to premature closure of the aperture. Ultimately, we ended up with six F814W images with an average exposure time of 1213~s. \flc\ images from both epochs were downloaded from the MAST archive.

\begin{figure}
    \centering
    \includegraphics[width=\columnwidth]{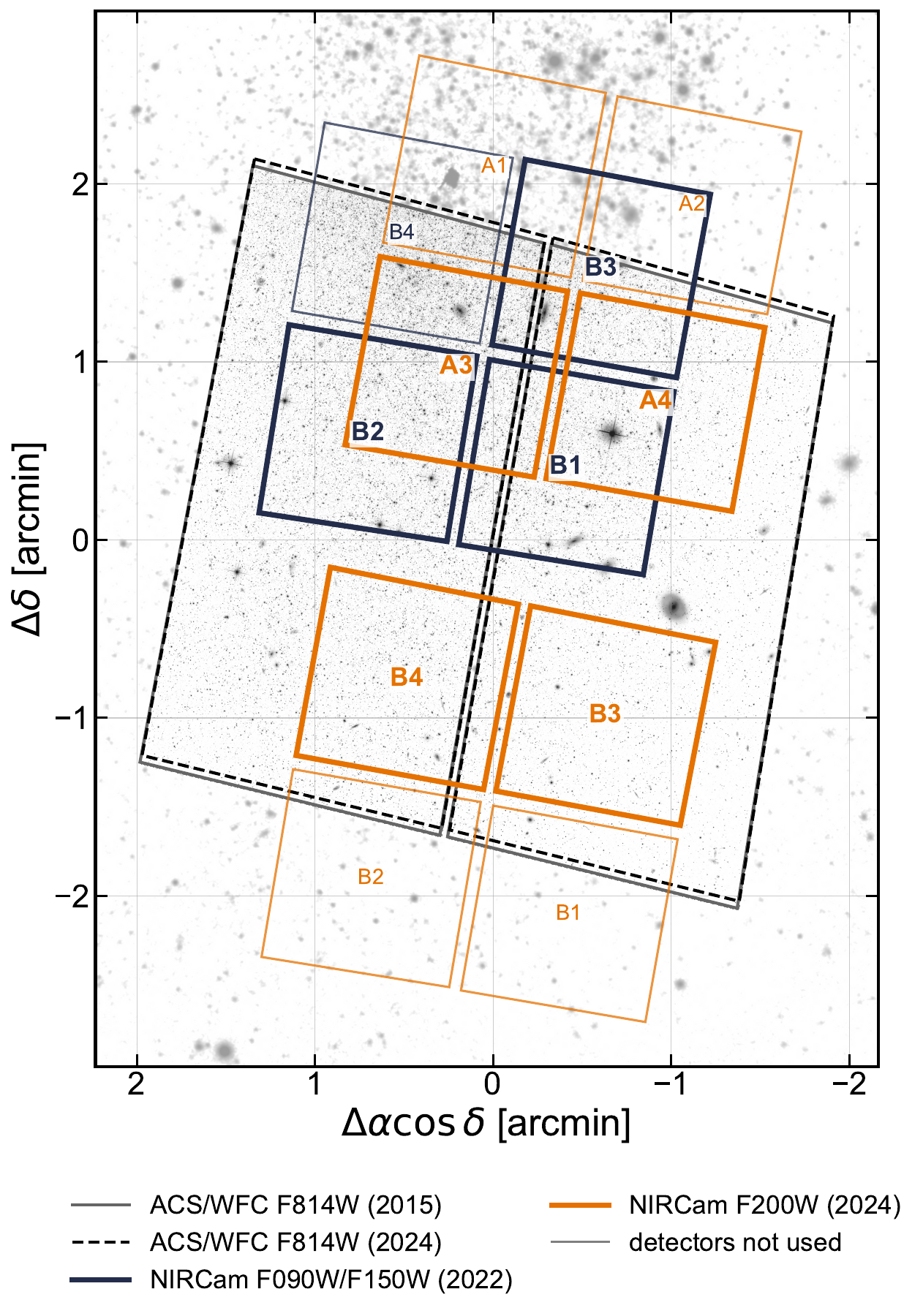}
    \caption{Relative placement of the ACS and NIRCam fields used for our PM analysis of WLM, shown on the sky (north up, east to the left). The offsets are relative to the center of the 2015 ACS field ($\alpha = 0\fdg4924$, $\delta = -15\fdg5226$). The background is the stacked 2015 ACS F814W image (``E1'') inside the 2015 ACS footprint and a Pan-STARRS1 $i$-band image \protect\citep{Chambers2016} elsewhere. The approximate footprints of the ACS observations (HST-GO 13768, PI D. Weisz, and HST-GO 17473, PI J. Warfield) are plotted as solid gray and dashed black outlines, respectively. The NIRCam footprint from the JWST Resolved Stellar Populations Early Release Science Program (PID 1334, PI D. Weisz; ``E2''), which provides the F090W and F150W images, is plotted in dark blue. The NIRCam footprint from JWST-GO 4570 (PI J. Warfield; ``E3''), which provides the F200W images, is plotted in orange. The NIRCam detectors are labeled, and those not used in any of our calculations are drawn with thinner, fainter outlines and unbolded labels.}
    \label{fig:wlm_footprint}
\end{figure}

\subsection{JWST NIRCam Imaging}

The NIRCam observations of WLM came from two programs. The first set were from the Resolved Stellar Populations Early Release Science Program (PID 1334, PI D. Weisz; \citealt{Weisz2023}) and were taken in 2022 July, forming a seven-year baseline with the E1 ACS observations; we refer to this set of imaging as ``E2.'' These observations used the F090W and F150W short-wavelength (SW) filters, paired with the F430M and F250M long-wavelength (LW) filters, and were designed to reach ${\rm SNR} \approx 10$ at ${\rm F090W} = 28.3$ and ${\rm F150W} = 28.1$ \citep{Weisz2023}.

The second set of NIRCam observations come from JWST-GO 4570 (PI J. Warfield), which imaged WLM in the F200W SW filter in 2024 July, forming a baseline of about nine years with E1. We chose F200W because, together with the E2 F090W and F150W imaging, it extends our second-epoch coverage across nearly the full SW wavelength range, and because its PSF is the best sampled of the SW filters. We also refer to this set of imaging (which was obtained as part of a joint program with the E3 HST data) as being a part of E3. These observations were designed to reach ${\rm SNR} \approx 23$ for F200W at the equivalent of ${\rm F090W} = 27$. For both programs, the NIRCam observations used the \texttt{MEDIUM8} read-out pattern with four dithered exposures per filter; the numbers of groups and integrations and the resulting effective exposure times are listed in Table~\ref{tab:obs}.

The E2 NIRCam field was centered on WLM to overlap the low-metallicity molecular clouds detected with ALMA and the deep archival optical HST imaging, with the HST ultraviolet (UV) and infrared (IR) imaging of WLM placed to overlap it in turn \citep{Weisz2023}. The overlap between the ACS and these NIRCam fields is limited to the B1 and B2 detectors and approximately half of the B3 and B4 detectors (Figure~\ref{fig:wlm_footprint}).
We do not use E2's B4 detector because its PMs in both F090W and F150W show large offsets from those of the other detectors (Section~\ref{ssec:res_overview}). We have not identified the cause. Wisps are most prominent on B4,\footnote{\url{https://jwst-docs.stsci.edu/known-issues/nircam-known-issues/nircam-scattered-light-artifacts}} but the offsets are spread across the part of the detector that overlaps the ACS field and do not correlate with the local background gradient, and the gradients we measure would shift the fits by $\lesssim$0.02~mas~yr$^{-1}$. B4 also has regions with a higher density of charge traps,\footnote{\url{https://jwst-docs.stsci.edu/jwst-near-infrared-camera/nircam-performance/nircam-persistence}} and in our E2 exposures the latent images that bright stars leave in the following exposure are two to three times stronger on B4 than on B2 and B3. We have not established that this persistence causes the offset.
The E3 observations were instead centered and oriented to maximize overlap with the ACS field, which we prioritized over overlap with the E2 NIRCam fields. The two NIRCam epochs therefore overlap only partially (Figure~\ref{fig:wlm_footprint}), and for F200W we use the A3, A4, B3, and B4 detectors.

For both E2 and E3 NIRCam observations, we downloaded the Stage~2 \texttt{cal} files, which are calibrated individual-exposure (averaged over an exposure's integrations), non-distortion-corrected images, the effective equivalent of \flc\ images for NIRCam \citep{Bushouse2025}.\footnote{See \url{https://jwst-docs.stsci.edu/jwst-science-calibration-pipeline} for a full description of the NIRCam pipeline science data products.}

\subsection{Preliminary Source Catalogs} \label{ssec:catalogs}

Catalogs of native pixel source positions were created for each \flc\ image from both ACS epochs by running \texttt{hst1pass}\footnote{\url{https://www.stsci.edu/~jayander/HST1PASS/}} \citep{Anderson2022}, a \texttt{FORTRAN}-based effective PSF (ePSF) fitting routine for HST images, which executes ePSF-fitting on \flc\ files at the sub-pixel level. We then corrected source centroids by applying the ACS pipeline geometric-distortion correction (GDC; \citealt{KozhurinaPlatais2015}).

For the NIRCam observations, we generated catalogs using the still in-development NIRCam module of \texttt{jwst1pass},\footnote{\url{https://www.stsci.edu/~jayander/JWST1PASS/}} an adaptation of \texttt{hst1pass} to do high-precision pixel-level ePSF astrometry with imaging data from JWST's NIRISS, MIRI, and NIRCam instruments (some of the development of this software is documented in \citealt{Libralato2023} for the ``officially'' released NIRISS module).

The PSFs and GDCs for \texttt{jwst1pass} were calibrated using an LMC starfield that was singled out for its ``perfect'' stellar density and properties for full-field JWST distortion calibration (identified by James Rhoads and fine-tuned by \citealt{DiazMiller2007}, according to \citealt{Anderson2021}). This field was observed with HST and cataloged in 2006 (PID 10753, PI R. Diaz-Miller) and 2017 (PID 14911, PI M. Fall). Subsequent releases of Gaia catalog data (in particular, Gaia EDR3 in 2020 and DR3 in 2022; \citealt{Gaia2021, Gaia2023}) added even more astrometric (position and PM) data for calibration. The planned calibration strategy using these data, and then the successful use of this field for astrometrically calibrating the NIRISS instrument, is described in \citet{Anderson2021} and \citet{Libralato2023}, respectively.
For the LW exposures, where PSFs and GDCs are not available for all filters, we use the F277W GDC for all three LW filters, and we measure the F430M catalogs with the F444W PSF.

\begin{figure}
\centering
\includegraphics[width=\columnwidth]{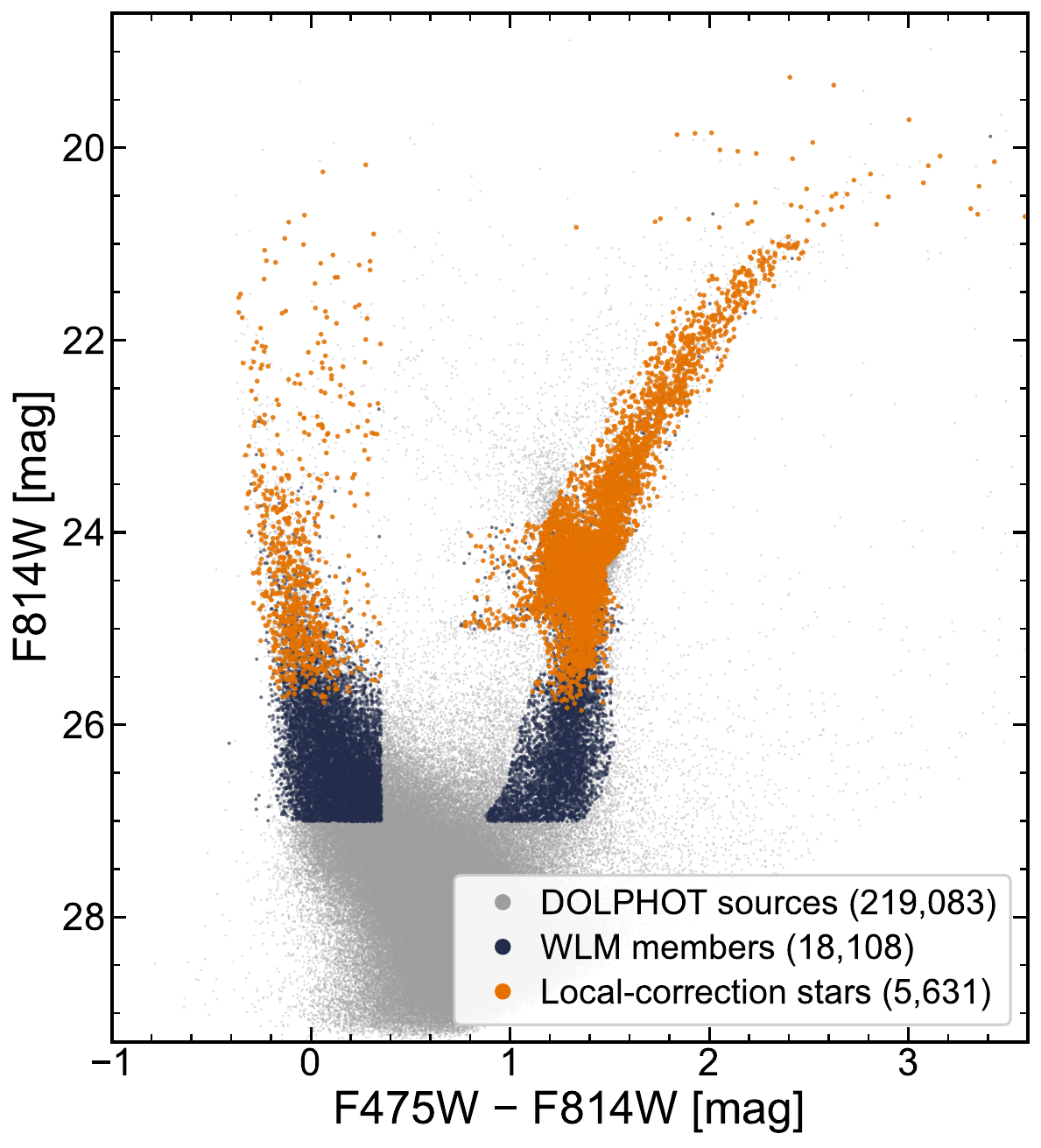}
\caption{Color--magnitude diagram of WLM from our first-epoch HST/ACS \texttt{DOLPHOT} photometry (Vega magnitudes). In gray, we plot all 219,083 \texttt{DOLPHOT} sources within the first-epoch ACS field that pass the catalog cuts (Section~\ref{ssec:catalogs}). The 18,108 stars selected as WLM members are plotted in blue, and the 5631 of them that serve as local-correction stars in the full-field HST-to-HST measurement (Section~\ref{ssec:localcorr}) are plotted in orange. The local-correction stars are all brighter than ${\rm F814W} = 25.9$, because the fainter members do not have template-fit positions in enough of the individual exposures. The local-correction stars of the HST-to-JWST measurements are not shown separately, because 97\% of them are among the stars plotted in orange.}
\label{fig:cmds}
\end{figure}
For the identification of likely WLM member stars, we used photometric catalogs generated with \texttt{DOLPHOT} \citep{Dolphin2000, Dolphin2016} from the first-epoch HST data (to ensure full-field coverage). These catalogs were generated by running \texttt{DOLPHOT} in the manner described by \citet{Williams2014} and \citet{Savino2022, Savino2023}, and were cut down to likely stellar sources using cuts in both F475W and F814W of: SNR~$\geq 4$, $|{\rm sharp}| \leq 0.458$, and ${\rm crowd} \leq 0.75$.\footnote{Definitions for these parameters can be found on the \texttt{DOLPHOT} user's manual webpage: \url{https://dolphot-jwst.readthedocs.io/en/latest/postprocessing/catalogs.html}.}
Within the first-epoch ACS field, this leaves 219,083 sources.
Membership was then selected from the CMD, using only sources with an F814W \texttt{DOLPHOT} flag of 0 and a defined color (i.e., a detection in both filters).
We selected as members the sources on the RGB, from its tip at ${\rm F814W} = 20.85$ down to ${\rm F814W} = 27$, that lie within 0.08--0.25~mag (2.5 times the rms in each 0.2~mag bin) of the median color of the sources with SNR~$\geq 10$, ${\rm crowd} \leq 0.20$, and $|{\rm sharp}| \leq 0.10$. We also selected the sources in three boxes on the CMD: the red clump, the asymptotic giant branch above the tip of the RGB, and the blue plume of upper main-sequence and blue helium-burning stars.\footnote{The red clump box is $23.9 \leq {\rm F814W} \leq 25.0$ and $0.75 \leq {\rm F475W}-{\rm F814W} \leq 1.35$, the asymptotic giant branch box is $19.0 \leq {\rm F814W} \leq 20.85$ and $1.3 \leq {\rm F475W}-{\rm F814W} \leq 4.0$, and the blue plume box is $20.0 \leq {\rm F814W} \leq 27.0$ and $-0.5 \leq {\rm F475W}-{\rm F814W} \leq 0.35$.}
We left out the region between the blue plume and the RGB brighter than the red clump, where red helium-burning stars are mixed with MW foreground dwarfs.
This gives 18,108 WLM members.

From these members, we draw the stars used to define the transformations between epochs (frame stars) and the stars used for the local correction (Section~\ref{ssec:localcorr}), using the F814W SNR, crowd, and sharp values. The frame stars---selected to be the brightest and most well-measured sources---are the members with SNR~$\geq 50$, ${\rm crowd} \leq 0.03$, $|{\rm sharp}| \leq 0.02$, and ${\rm F814W} \leq 26$, which gives 3702 frame stars. The local-correction stars are members with SNR~$\geq 10$, ${\rm crowd} \leq 0.20$, $|{\rm sharp}| \leq 0.10$, and ${\rm F814W} \leq 27$ (14,720 stars), of which we use the 14,531 that are more than 0\farcs4 from any of our background galaxies. In each run, a star is then used only if it has a template-fit position in enough of the individual exposures and moves with the bulk of WLM (Sections~\ref{ssec:comoving} and \ref{ssec:localcorr}).
We then crossmatched these \texttt{DOLPHOT} catalogs with our \texttt{hst1pass} and \texttt{jwst1pass} catalogs, giving us independent lists of stellar pixel positions for each HST exposure and each NIRCam detector that we use. We used the \texttt{DOLPHOT} catalogs for the membership selection because they were already in hand and provide somewhat deeper photometry, and because requiring a star to appear in both the \texttt{DOLPHOT} and the \texttt{hst1pass} or \texttt{jwst1pass} catalogs is an additional check that it is real. In principle, the same selection could be made from the \texttt{hst1pass} photometry alone. We did not apply any additional selection using the NIRCam photometry. In Figure~\ref{fig:cmds}, we plot the CMD of WLM with the selected members and with the stars that end up as local-correction stars in the full-field HST-to-HST measurement.

\begin{figure*}
    \centering
    \includegraphics[width=\textwidth]{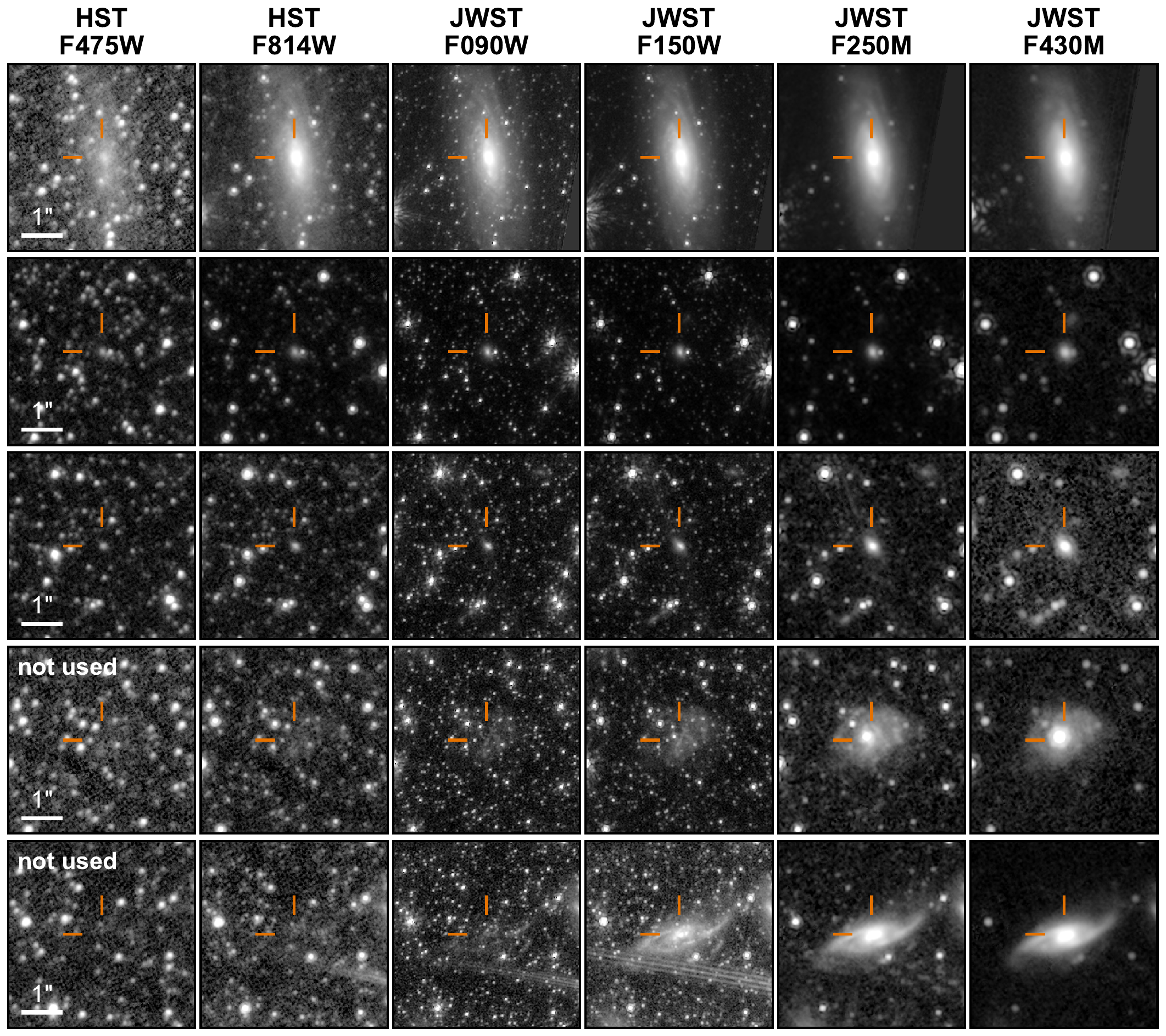}
    \caption{Background galaxies as they appear in two ACS/WFC filters from the first epoch (F475W and F814W) and in four NIRCam filters from E2 (SW F090W and F150W, LW F250M and F430M). The top three rows show galaxies used in our F090W measurement: from top to bottom, one of the brightest, one of about median brightness, and one of the faintest. The bottom two rows show galaxies that we do not use in any measurement, as extreme cases: they are nearly invisible in the ACS filters, but prominent in the LW filters (the streaks across the bottom row are diffraction spikes of a nearby bright star). Each panel is $5''$ on a side, with north up and east to the left, and is scaled independently; the orange ticks point to the galaxy. The galaxies stand out more and more from WLM's stars toward longer wavelengths, which is why we use the LW images to identify them, whereas their positions are measured in F814W and in the SW filters.}
    \label{fig:galsbywavelength}
\end{figure*}

As in \citet{Warfield2026}, we constructed our background galaxy catalog by eye in SAO~DS9 \citep{SAO2000}. The first pass was done with the stacked E1 ACS images and then additional passes were made for the footprint of the E2 field using the NIRCam LW F430M images, in which the galaxies are much easier to identify (Figure~\ref{fig:galsbywavelength}). We ended up with a list of 331~galaxies across the two ACS detectors. We also tested selecting galaxies automatically by color \citep{Warfield2023b}, but this split extended galaxies into multiple sources, picked up stars superimposed on galaxies, and missed many of the best extended galaxies, so that cleaning the automated list took longer than building it by eye.

\section{Methodology} \label{sec:methods}

\subsection{Overview} \label{ssec:overview}

We measure PMs using the background-galaxy template-fitting procedure proposed by \citet{Mahmud2008}, first applied by \citet{Sohn2012, Sohn2013}, and described step by step in \citet{Warfield2026}. In brief, we:
\begin{enumerate}
    \item Build a master list of stars and a twice-supersampled, distortion-corrected stack from the first-epoch exposures;
    \item Find the transformation between every later exposure and the master frame;
    \item Build a convolution kernel for each later exposure that accounts for the change in PSF between epochs;
    \item Measure the position of every star and background galaxy in every exposure of both epochs by fitting it with its template from the stack, distorted, and convolved with the kernel for the later exposures;
    \item Transform every exposure into the comoving frame of the target's stars, in which each background galaxy's change in mean position between epochs measures the reflex of the target's systemic PM; and
    \item Correct each galaxy for residual, spatially-correlated offsets using nearby member stars, and combine the galaxies into a systemic PM.
\end{enumerate}
The HST-to-HST measurement follows these steps as in \citet{Warfield2026}, and the HST-to-JWST measurement uses the same steps with NIRCam exposures in place of the second-epoch ACS exposures.

Most of what changes for NIRCam enters in step~2, the registration of the NIRCam exposures to the HST master frame (Section~\ref{ssec:register}), and step~3, the kernel that matches the resolution of the two epochs (Section~\ref{ssec:kernel}).
However, we have also changed several steps relative to \citet{Warfield2026} for both the ACS and NIRCam second epochs. These are:
\begin{enumerate}
    \item[a.] We measure the positions of stars with the same template fits as the galaxies, starting each fit at the star's \texttt{hst1pass} or \texttt{jwst1pass} position (Section~\ref{ssec:fits});
    \item[b.] We draw the frame transformation and local-correction stars from WLM members identified from the CMD (Section~\ref{ssec:catalogs});
    \item[c.] We add cuts on the quality of each galaxy's template fit and on the fraction of starlight in its fit stamp (Sections~\ref{ssec:crowding} and \ref{ssec:combine}); and
    \item[d.] We correct the HST-to-JWST PMs for detector-dependent biases measured from injection tests (Section~\ref{ssec:injection}).
\end{enumerate}

In contrast to ACS's \flc\ images, which are multi-extension and have consolidated GDCs associated with them that correct both WFC detectors at the same time, there are separate \cal\ files and associated GDCs for each of the NIRCam detectors, and the precise offsets between NIRCam detectors are not as well documented as the relationship between ACS's WFC detectors. Therefore, for cases of precision astrometry, it is ideal to process each detector separately. We treat each of the 14 NIRCam detector and filter combinations that we use (B1--B3 in F090W and F150W, and A3, A4, B3, and B4 in F200W; and, in the LW channel, B5 in F250M and F430M, and A5 and B5 in F360M) as a separate unit throughout, alongside the full HST field.

In the subsections below, we go through the pipeline chronologically and detail changes that are required to be made from the HST-to-HST process in \citet{Warfield2026}.

\subsection{A Common First-epoch Frame} \label{ssec:frame}

Following \citet{Warfield2026}, we first cross-match the \texttt{hst1pass} catalogs of the 26 first-epoch F814W exposures to build a master list of stars. To do this, we use the \texttt{xym2mat} and \texttt{xym2bar} routines.\footnote{\texttt{xym2mat}, \texttt{xym2bar}, and \texttt{img2stk} (used below) are available at \url{https://www.stsci.edu/~jayander/FORTRAN/}.}
We use the bright, well-measured sources in each catalog (\texttt{hst1pass} quality of fit $q \leq 0.05$) and keep the 1172 stars found in at least 20 of the 26 exposures.
Unlike \citet{Warfield2026} for Draco~II, we do not restrict this list to CMD members because WLM's stars overwhelmingly dominate the field.

The master frame is the distortion-corrected frame of the first exposure and each star's master position is the average of its positions in all of the first-epoch exposures after transforming each exposure into this frame with a six-dimensional linear transformation (6DLT).\footnote{$X_{\rm trans} = A_X + B_X X + C_X Y$; $Y_{\rm trans} = A_Y + B_Y X + C_Y Y$.}
We iterate the matches and transformations until they converge. In each pass, \texttt{xym2mat} fits the 6DLT to the matched stars and discards the matches that lie farther than a set tolerance from it, with the tolerance shrinking to 0.05~pixels over the passes, and \texttt{xym2bar} takes each star's master position to be the mean of its transformed positions, after rejecting, one at a time, positions that lie both more than 0.03~pixels and more than twice the rms of the others from their mean. At convergence, the 6DLT for each exposure is determined from 999--1072 stars, with rms residuals of 0.008--0.010~pixels (0.4--0.5~mas) per coordinate. Using these transformations, we then build a twice-supersampled (25~mas/pixel), distortion-corrected stack of the 26 exposures with \texttt{img2stk}, in the same fashion as Section~3.4 of \cite{Sohn2012}. The templates for every star and galaxy (Section~\ref{ssec:fits}) and the reference image for every kernel (Section~\ref{ssec:kernel}) are cut from this stack.

\subsection{Registering the Second-epoch Exposures} \label{ssec:register}

Because the E3 HST observations were taken with the same pointing and orientation as was E1, \texttt{xym2mat} can match each exposure's stars to the master list directly in detector coordinates, which already lie close to the same values pre-transformation. This gives 933--972 matched stars per exposure, with rms 6DLT residuals of 0.015--0.016~pixels (0.8~mas) per coordinate.

This is not the case for the cross-observatory methodology. The NIRCam exposures differ from the ACS master frame in pixel scale ($\sim$31 versus 50~mas/pixel), orientation (by $\sim$3--5$^\circ$ in our fields), and distortion, so \texttt{xym2mat}'s own pattern search finds no matches between them.
We therefore start from the sky coordinates that \texttt{hst1pass} and \texttt{jwst1pass} provide. We first fit the transformation from master positions to sky coordinates using the first-epoch matches, which is linear to within $\sim$2~mas because the master frame is distortion-corrected. For each NIRCam exposure, we then match its stars to the master stars within 1$''$ on the sky, remove the median offset between the two observatories' World Coordinate System (WCS) solutions (typically a few tenths of an arcsecond), and rematch within 0\farcs1.
A single nearest-neighbor match within a fixed radius (e.g., $\sim$0\farcs5) can fail when the WCS offset is comparable to that radius, which removing the median offset avoids. These matches (a few tens per exposure) seed \texttt{xym2mat}, which then solves for the 6DLT between the exposure and the master frame and iteratively rematches the stars in pixel space. After two passes of \texttt{xym2mat} and \texttt{xym2bar}, each NIRCam SW exposure is matched to the master list by 68--183 stars (fewer than for the ACS exposures because each NIRCam detector covers only part of the ACS field; Figure~\ref{fig:wlm_footprint}), with rms 6DLT residuals of 0.007--0.016 master pixels (0.3--0.8~mas) per coordinate, and each LW exposure by 141--443 stars, with rms residuals of 0.019--0.021 master pixels (0.9--1.1~mas).

\subsection{Matching the Resolution of the Two Epochs} \label{ssec:kernel}

\begin{figure*}
\centering
\includegraphics[width=\textwidth]{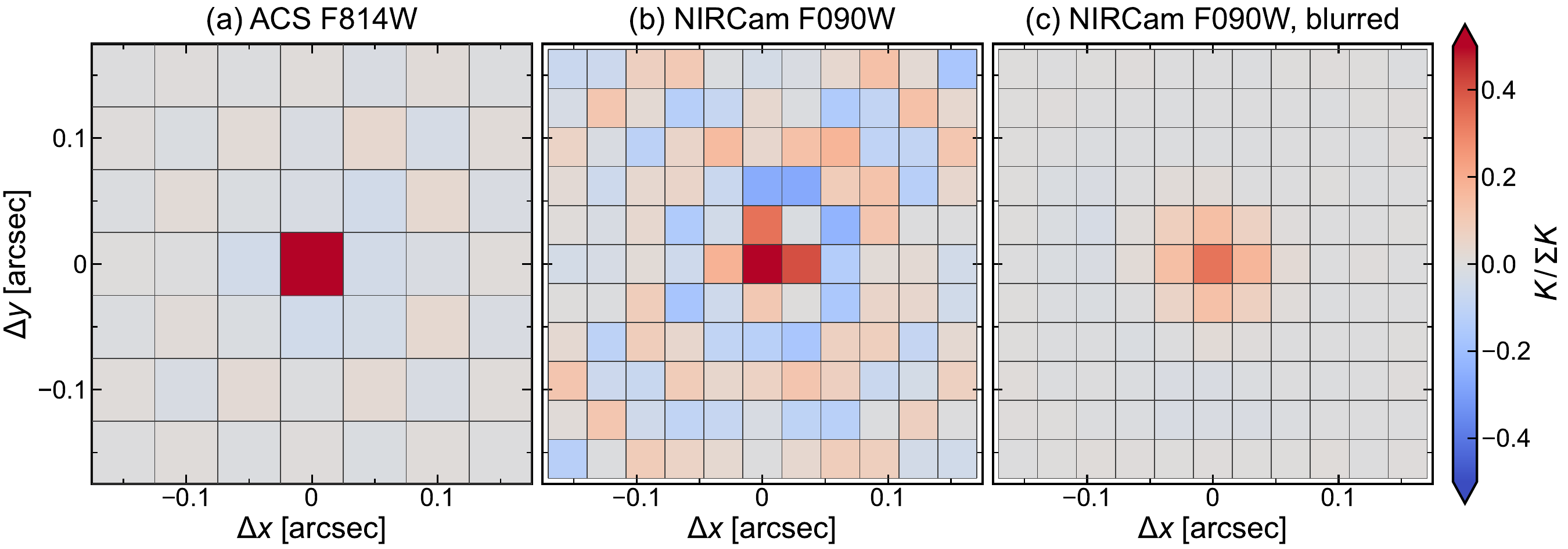}
\caption{Convolution kernels relating the stacked first-epoch HST/ACS F814W image of WLM to individual later-epoch exposures, each divided by its sum and shown on a common color scale that saturates at $\pm0.5$. (a) A second-epoch HST/ACS F814W exposure. (b) A JWST/NIRCam B2 F090W exposure on the original, unblurred image, for which the kernel would need to sharpen the stacked image. (c) The same NIRCam exposure as in (b) after blurring it to a stellar FWHM of 105~mas. The peak values of the three kernels are 1.18, 0.82, and 0.33, and their negative pixels sum to $-0.48$, $-4.60$, and $-0.44$, respectively. The ACS kernel is $7\times7$ ACS pixels and the NIRCam kernels are $11\times11$ NIRCam pixels, so all three cover $\sim$0.34$''$ on the sky.}
\label{fig:kernels}
\end{figure*}

In order to measure a background galaxy's position in a second-epoch exposure, its template, which is cut from the stacked first-epoch image, first has to be made to look like the galaxy as it appears in that exposure. For two epochs of HST imaging, this is done with a convolution kernel built from bright, isolated stars, fit such that the stacked image convolved with the kernel best matches each second-epoch exposure. In effect, this is a data-driven comparison of the PSFs of the two epochs, which accounts for centroid shifts owing to changes in the PSF between epochs (which can differ between point sources and extended sources) as well as changes in the shapes of the sources themselves \citep{Sohn2012}. As in \citet{Sohn2012} and \citet{Warfield2026}, we use $7\times7$~pixel kernels for the HST E3 images (fit here to $9\times9$~pixel stamps around each star; Figure~\ref{fig:kernels}a).
For NIRCam, we instead use $11\times11$~pixel kernels, which are scaled from the HST case to roughly cover the same area of the sky ($\sim$0.34$''$) at NIRCam's pixel scale.\footnote{In our tests on blurred NIRCam images, the outermost ring of a $7\times7$~pixel kernel sums to $-0.36$ to $-0.46$ times the kernel's total, an artificial negative ring, compared to $-0.02$ to $+0.08$ for an $11\times11$~pixel kernel.}

However, this only works when the later-epoch exposures are no sharper than the stacked image, since the stacked image contains essentially no information on scales finer than its own PSF, and no convolution can put that information back. Because NIRCam's SW image PSFs are considerably sharper than ACS F814W's (stellar FWHMs of $\sim$45, 55, and 67~mas in F090W, F150W, and F200W, against $\sim$87~mas in a single F814W exposure), a kernel that matches the stacked image to these exposures would need to sharpen it, adding detail that is simply not present in the HST data.
Rather than failing outright, the fit compensates with large positive and negative values scattered across the kernel (Figure~\ref{fig:kernels}b), so that the negative pixels of the kernel sum to $\sim$3--7 times the kernel's total, against $\sim$0.5 for both the HST-to-HST kernels and kernels fit to blurred NIRCam images (Figure~\ref{fig:kernels}c). This is somewhat less severe in F200W ($\sim$3--4 times), where the NIRCam PSF is closer to that of the stacked image.
Convolving a template with such a kernel adds noise and distorts the template rather than matching it to the data. Measuring the galaxy positions with these kernels makes the combined F090W PM uncertainties $\sim$1.6 times larger than with blurred images, and moves the combined $\mu_\delta$ values by $\sim$0.05--0.09~mas~yr$^{-1}$.

To get around this, we tested two approaches. In the first, we blur each NIRCam exposure (by convolving it with a Gaussian, as described below) to a stellar FWHM of 105~mas, chosen to be safely broader than the $\sim$90~mas stacked HST image, so that the kernel only ever has to broaden the template, and then fit kernels to the blurred images in the usual way. In the second, which we refer to as the ``reverse'' kernel, we instead fit a kernel that degrades each NIRCam exposure to match the stacked image, and then fit the unconvolved template to the convolved exposure. Both approaches work; the combined PMs from the two approaches agree within 0.07~mas~yr$^{-1}$ in F090W and F150W, but differ by 0.08 and 0.14~mas~yr$^{-1}$ (in $\mu_\alpha^*$ and $\mu_\delta$) in F200W. Because the blurred images show smaller biases in our injection tests ($\lesssim$0.02~mas~yr$^{-1}$, against $+0.01$ to $+0.04$~mas~yr$^{-1}$ for the reverse kernel), we adopt the blurred images for our final measurements.

To create these de-sharpened, blurred images, we convolve each NIRCam exposure with a round Gaussian, adjusting the width of the Gaussian until bright, isolated stars in the blurred exposure measure 105~mas.
Ultimately, we used Gaussians with FWHMs of $\sim$86, 80, and 74~mas to convolve F090W, F150W, and F200W, respectively, and achieved blurred stellar FWHMs of 105.1--105.4~mas in every exposure.
We use a normalized convolution, so that pixels without valid data (e.g., bad or saturated pixels) do not contribute to their neighbors, and we mask any pixel for which more than 1\% of the Gaussian's weight falls on such pixels; masked pixels are given zero weight in the template fits (Section~\ref{ssec:fits}). Because the Gaussian is symmetric, blurring does not move the centroids of sources and leaves each exposure's WCS unchanged, so the transformations solved for in Section~\ref{ssec:register}, which are measured from the \texttt{jwst1pass} catalogs of the original images, work for the blurred images as well.

For the LW units, we use $7\times7$~pixel kernels, which cover $\sim$0.44$''$ at the 63~mas LW pixel scale. We blur the F250M exposures to 105~mas in the same way, but the F360M (128--129~mas) and F430M stellar FWHMs are already broader than the stacked image, and so we do not blur them.

Following \citet{Sohn2012}, we also fix the position of each kernel star to its position as measured by \texttt{hst1pass} or \texttt{jwst1pass}, and solve for the kernel with only the constraint that it sums to one. This removes the degeneracy between shifting the kernel and shifting the scene, and it means that the kernel itself carries any systematic offset between where the template fit and the PSF fit place the center of a star.
As a result, the template-fit positions of stars measured with these kernels agree with their \texttt{jwst1pass} positions to $\sim$0.02~pixels on average, where they were offset by up to $\sim$0.4~pixels when the kernel's shift was instead fit for each star.
This matters more for the kernels than for the position fits. Errors of a fraction of a pixel in the registrations of the kernel stars distort the kernel's shape, and the kernel imparts this distortion on the convolved galaxy templates.

\subsection{Measuring Positions with Template Fits} \label{ssec:fits}

We measure the position of every star and background galaxy in every exposure of both epochs with the same template fitting procedure, following steps~10 and 11 for ``Method 2'' in Section~4 of \citet{Warfield2026}.
For each source, the template is taken from the stacked first-epoch image (Section~\ref{ssec:frame}), mapped onto the exposure's pixels through the exposure's 6DLT and the inverse of its GDC, and, for the later epochs, convolved with the exposure's kernel (Section~\ref{ssec:kernel}).
The template is then shifted against the exposure on a grid of sub-pixel offsets (in steps of 0.1~pixels out to $\pm 1$~pixel, and then in steps of 0.01~pixels around the best of these). The total flux of the template cut-out is scaled to that of the convolved individual-exposure image, and the offset that minimizes the sum of the absolute residuals over a $5\times5$~pixel stamp of the exposure is recorded as the source's position.
Both the template and the stamp are sky-subtracted, with the sky measured in annuli of 0\farcs30--0\farcs50 around the source in the stack and 6--10~pixels in the exposure. We keep the sum of the absolute residuals, divided by the flux in the stamp, as the quality of fit, $q$, of each measurement.
Unlike \citet{Warfield2026}, we give zero weight in the fit to pixels that are flagged as unusable in the exposure's data-quality array (e.g., saturated pixels, hot pixels, and cosmic rays), and we discard measurements that fail, fall off the image, reach the edge of the search grid, or lie on a satellite trail or across the ACS chip gap. In the HST data of Draco~II used by \citet{Warfield2026}, giving these pixels zero weight lowers the 90th-percentile scatter of a source's positions between the exposures of the first epoch from 0.16 to 0.10~pixels for stars and from 0.54 to 0.21~pixels for galaxies; 6--18\% of the fit stamps in each exposure contain a pixel flagged as a cosmic ray.

As described in Section~\ref{ssec:overview}, we measure the stars with these same template fits, rather than adopting their \texttt{hst1pass} positions as in \citet{Warfield2026}. This follows \citet{Sohn2012, Sohn2013, Sohn2017, Sohn2020}: template-fit positions of stars are less precise than their PSF-fit positions, but measuring the stars and the galaxies in the same way means that a systematic in the templates or kernels that is common to both cancels in the comoving frame. For the stars, we also change where each fit starts. Rather than starting at the star's predicted position in the exposure, we start each star's fit at its \texttt{hst1pass} or \texttt{jwst1pass} position in that exposure, without recentering its template, and the fit is then free to move from there; a star with no \texttt{hst1pass} or \texttt{jwst1pass} match in an exposure is not measured in that exposure.
We do this because the PSF fits of \texttt{hst1pass} and \texttt{jwst1pass} already find the centers of stars to well within our required astrometric precision, even in crowded fields, so there is no need to recenter the star templates, which would only place each template on its brightest stack pixel. For isolated stars, the difference is mostly negligible, but in a field as crowded as the center of WLM,\footnote{The \texttt{DOLPHOT} catalog for E2's B2 detector contains an average $\sim 7.6$~stars per arcsec$^2$, compared to $\sim 0.19$~stars per arcsec$^2$ on the most crowded detector for this ERS program's NIRCam observations of Draco~II.} a brighter neighbor has a real chance of biasing that choice. Starting from these positions, together with kernels built around the same positions (Section~\ref{ssec:kernel}), also keeps the template-fit positions of the stars on the same system as the \texttt{hst1pass} and \texttt{jwst1pass} catalogs from which the registrations of Sections~\ref{ssec:frame} and \ref{ssec:register} are measured.

For the background galaxies, we use the 331 galaxies of our WLM list (Section~\ref{ssec:catalogs}), and, as in \citet{Warfield2026}, we center each template on the brightest stack pixel within 5~stack pixels (0\farcs125) of the galaxy's listed position. The fit of a galaxy has a problem that the fit of a star does not. The $5\times5$~pixel stamp is centered on the pixel nearest the galaxy's predicted position in the exposure, and because the 6DLTs follow WLM's stars, this is where the galaxy would be if it moved with them. A galaxy's reflex motion moves it away from this position, and when a galaxy is faint and its light is truncated by the stamp, the fit is pulled back toward the center of the stamp, shrinking the measured motion. We therefore refit the galaxies with an iterated prediction: after the first fit, we shift each galaxy's predicted position in every exposure of an epoch by its median measured offset from that prediction over the epoch's exposures, and fit again.
We repeat this four times, after which the offsets of most galaxies no longer change. For a small fraction of the measurements (3\% of those in the first-epoch ACS exposures, 11\% in the second-epoch ACS exposures, and 3--6\% in the NIRCam exposures we tested), the fitted position instead alternates between two neighboring stamps from one pass to the next.
Averaging the positions from the last two passes changes our PMs for the full HST field and the B2 detector by at most 0.007~mas~yr$^{-1}$, so we keep the positions from the fourth pass. The stars do not need this step, since the 6DLTs already follow their motions and their fits start at their measured positions.

For WLM, whose reflex motion amounts to $\lesssim 1$~mas over our baselines, this magnitude-dependent effect is small, but it is easily measured in the HST data of Draco~II used by \citet{Warfield2026}, whose reflex motion is $\sim 1.4$~mas~yr$^{-1}$. In those data, before the refit, the PMs of the faintest quarter of the galaxies came out at only $\sim$61\% of the PMs of the brightest quarter (along the direction of the bright galaxies' mean motion). After switching to using this refit, this improves to $\sim$95\%.
It also matters more for the HST-to-JWST measurements than it otherwise would: because faint galaxies are easier to identify in the NIRCam images (Figure~\ref{fig:galsbywavelength}), our lists go somewhat deeper than they would from the HST images alone, and so the faint galaxies can make up a significant part of a measurement. In \citet{Warfield2026}, the cut on the total PM uncertainty of each galaxy ($<1$~mas~yr$^{-1}$) likely limited this effect, in part, by removing many of the fainter galaxies, which also tend to have the largest uncertainties. With the refit, the faint galaxies can carry their share of the measurement, where before we relied heavily on the brightest galaxies.

Because the stamp and the sky annulus are defined in exposure pixels, they cover different amounts of sky in the two instruments: a NIRCam SW stamp is 0\farcs155 on a side, against 0\farcs25 for ACS, and its sky annulus (0\farcs19--0\farcs31) lies closer to the galaxy. We tested larger stamps ($7\times7$ and $9\times9$~pixels) and a sky annulus matched on the sky to the ACS annulus (10--16~NIRCam pixels) on the B2 F090W and F150W exposures. None of these is clearly better than the $5\times5$~pixel stamp and the 6--10~pixel annulus, as judged by how well the PMs of the individual galaxies agree with those from the HST-to-HST measurement of the same area, and so we keep them. However, repeating the final F090W measurement with a $7\times7$~pixel stamp changes the PMs of the individual galaxies by $\sim$0.2~mas~yr$^{-1}$ (rms), comparable to their uncertainties, and moves the combined F090W PM by $(-0.04, +0.07)$~mas~yr$^{-1}$, which is comparable to its statistical uncertainty.

\subsection{Crowding in the Central Fields of WLM} \label{ssec:crowding}

\begin{figure}
\centering
\includegraphics[width=\columnwidth]{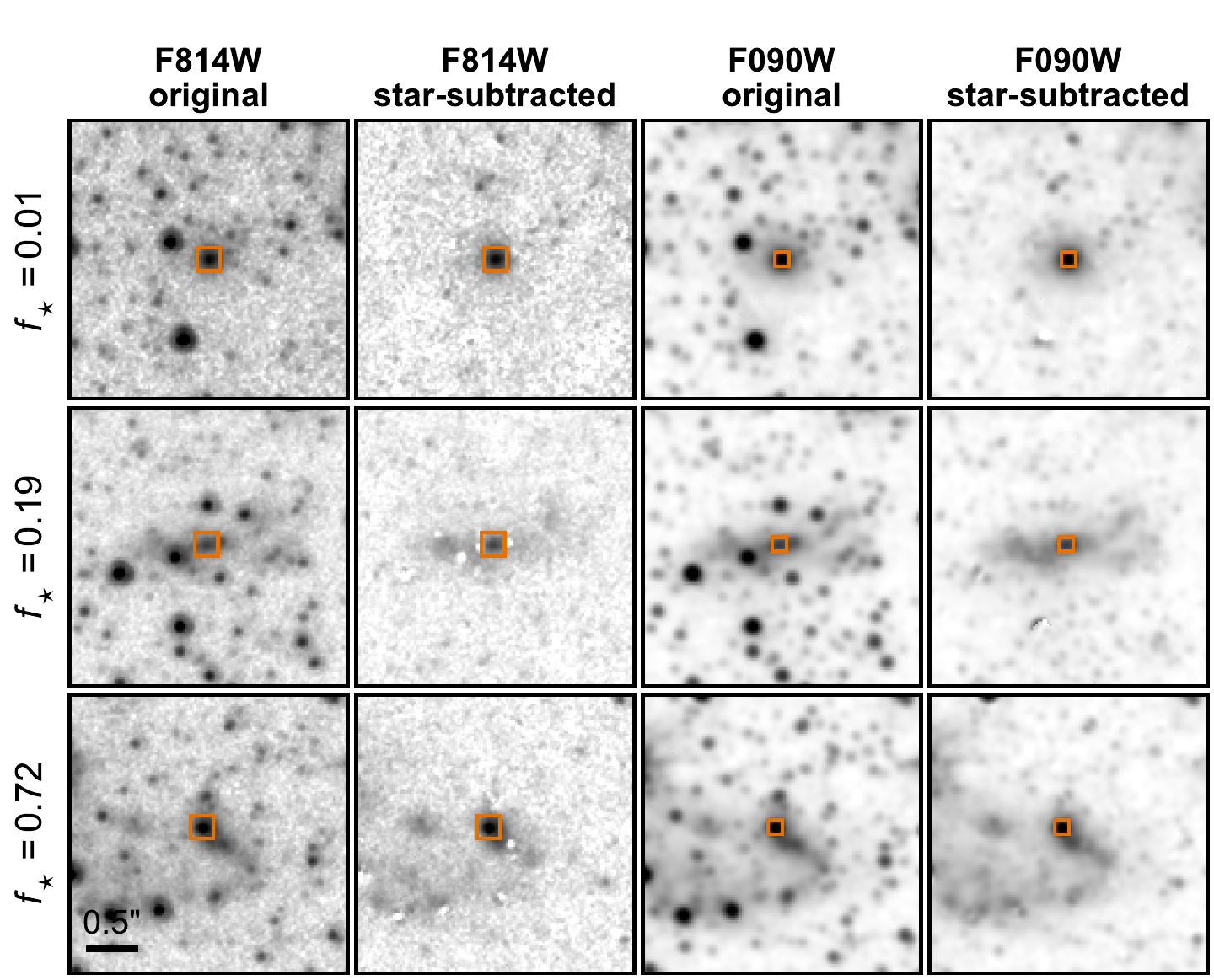}
\caption{Three background galaxies in the central E2 B2 field (one per row), in $3'' \times 3''$ cutouts, ordered by the stellar fraction of their fit stamps, $f_\star$ (Section~\ref{ssec:crowding}); the galaxy in the bottom row exceeds our limit of $f_\star < 0.5$ and is not used. From left to right: the stacked first-epoch ACS/WFC F814W image; the same stack built from star-subtracted exposures; the four E2 NIRCam F090W B2 exposures, blurred to a stellar FWHM of 105~mas as in our fits (Section~\ref{ssec:kernel}), distortion-corrected, resampled to the same grid and orientation, and averaged; and the same from the star-subtracted, blurred F090W exposures. The orange squares outline the $5\times5$~pixel fit stamps ($0\farcs25$ on a side for ACS and $0\farcs155$ for NIRCam). Within each row, the two F814W panels share one display stretch, as do the two F090W panels. In each, several WLM stars, some nearly as bright as the galaxy's core, fall on and around the galaxy. The white spots in the subtracted panels are subtraction residuals, which are given zero weight in the fits.}
\label{fig:crowding}
\end{figure}

Imagine modeling the light of a source as a circular Gaussian profile, $f$, with total flux $F$ and width $\sigma$.
For our methodology, what matters is not only the flux, but how steeply the profile changes across the fit stamp. In a least-squares fit for a shift, $s$, with uniform background noise, $\chi^2$ grows for small $s$ as $\Delta\chi^2 \propto s^2 \int (\partial f/\partial x)^2\,d^2x$, so the constraint that each source places on the shift is set by its image gradient power,
\begin{equation}
    P_{\rm grad}=\int |\nabla f|^2 \, d^2x = \frac{F^2}{4 \pi \sigma^4}.
\end{equation}
$P_{\rm grad}$ is larger for sources with smaller widths.
This is somewhat intuitive: If we had a true point source with $\sigma=0$ located in the middle of a $5\times5$~pixel postage stamp with no background, any whole-pixel shift, $s$, will maximize the residual across that stamp. For a Gaussian, the residual instead grows gradually, as $\Delta\chi^2 \propto s^2 F^2/\sigma^4$ for $s \ll \sigma$. So, at a fixed flux, a source twice as wide must be shifted four times as far to produce the same residual.

For real data, point (stellar) sources will have tighter profiles, with $\sigma \approx 0.8$~pixels for a star on the ACS detector.
Background galaxies are more extended, with $\sigma \approx 3$~pixels for a fairly compact background galaxy. The ratio of the gradient power for the two is:
\begin{equation}
    \frac{P_{\rm star}}{P_{\rm gal}} = \frac{\sigma_{\rm gal}^4}{\sigma_{\rm star}^4} \frac{F_{\rm star}^2}{F_{\rm gal}^2} \approx 200 \frac{F_{\rm star}^2}{F_{\rm gal}^2}.
\end{equation}
Therefore, a star only needs to have $\sim7\%$ the flux of a background galaxy in order to match or exceed its gradient power in the fit.

When a star and a galaxy share a stamp, the template contains both, but we fit only a single shift. To first order, and when the two profiles do not overlap much, the fitted shift is then the gradient-power-weighted mean of the shifts of the two sources,
\begin{equation}
    s_{\rm fit} \approx \frac{P_{\rm gal}\, s_{\rm gal} + P_{\rm star}\, s_{\rm star}}{P_{\rm gal} + P_{\rm star}}. \label{eq:sfit}
\end{equation}
These estimates assume Gaussian profiles and uniform background noise, and only the part of a star's light that falls inside the stamp counts. Real galaxies, with cores and knots, have more gradient power than a Gaussian of the same overall width, so $\sigma_{\rm gal}$ here is an effective width.

Many of our background galaxies have WLM stars close-by or even overlapping them (Figure~\ref{fig:crowding}, first and third columns), particularly in the central fields covered by E2's B1 and B2 and by E3's A3 and A4. Since stars are at rest in the comoving frame ($s_{\rm star} = 0$ in Equation~\ref{eq:sfit}), they scale the measured motion of the galaxy by $P_{\rm gal}/(P_{\rm gal} + P_{\rm star})$, pulling it toward zero motion. 

In the hope of reducing this pull, we tested subtracting the known stars from the images in order to reduce the amount of stellar flux that falls within each galaxy's $5\times5$ fit stamp.
In each exposure of both epochs, we fit and subtracted the sources of our \texttt{DOLPHOT} catalog with \texttt{hst1pass} or \texttt{jwst1pass} (except for those within 5~pixels of a listed galaxy in order to avoid subtracting flux from bright galaxy cores), built the template stack from the subtracted first-epoch exposures (Figure~\ref{fig:crowding}, second column), and gave zero weight in the galaxy fits to the pixels nearest each subtracted star, where its residuals are largest.
However, the results from the subtracted images were not noticeably better (in terms of agreeing with Gaia literature PM values), and subtraction moves the HST-to-JWST and HST-to-HST PMs over the same areas in the same direction (by about $-0.04$~mas~yr$^{-1}$ in $\mu_\delta$) leaving the comparison between the two observatories unchanged. Additionally, using star-subtracted images leaves 4--14 fewer galaxies in each combined table. In our injection tests (Section~\ref{ssec:injection}), subtraction recovers only part of the motion that is lost to crowding (a gain of $\sim$0.01~mas~yr$^{-1}$ at WLM's PM). We therefore do not use star subtraction for our final measurements.

Instead, we remove the galaxies whose fit stamps are dominated by starlight (as in the bottom row of Figure~\ref{fig:crowding}). For each galaxy, we estimate the stellar fraction of its stamp, $f_\star$: the flux that the sources of our \texttt{DOLPHOT} catalog lying more than 1~pixel from the galaxy's center place inside the stamp (from their F814W magnitudes and an empirical PSF built from the stacked first-epoch image), divided by the total sky-subtracted flux of the stamp in that image. In the full-field HST-to-HST measurement, the PM is stable when only galaxies with $f_\star$ below $\sim$0.3 are used, and shrinks toward zero as more crowded galaxies are added: from $(+0.080~\pm~0.017,\allowbreak -0.070~\pm~0.016)$~mas~yr$^{-1}$ for $f_\star < 0.3$ (116 galaxies) and $(+0.075~\pm~0.016,\allowbreak -0.068~\pm~0.014)$~mas~yr$^{-1}$ for $f_\star < 0.5$ (145 galaxies) to $(+0.052~\pm~0.015,\allowbreak -0.048~\pm~0.014)$~mas~yr$^{-1}$ for all 167 galaxies. We therefore keep only galaxies with $f_\star < 0.5$ (Section~\ref{ssec:combine}). The stricter threshold changes the full-field PM by $\leq$0.005~mas~yr$^{-1}$, but removes about a third of the galaxies on the NIRCam detectors. However, $f_\star$ is an upper limit on the stellar fraction of a stamp and not a measurement of it, because \texttt{DOLPHOT} also detects parts of the galaxies themselves as separate sources. On average, the catalog places about three times more flux in the stamps of our galaxies than in stamps at random positions in the field, and most of this excess comes from sources that \texttt{DOLPHOT} does not measure as clean point sources. The threshold of 0.5 is therefore an empirical choice. We test how the stars in a stamp change a galaxy's measured motion with injected galaxies in Section~\ref{ssec:injection}.

\subsection{The Comoving Frame of WLM} \label{ssec:comoving}

With every source measured in every exposure, we use HubPUG \citep{Warfield2023} to transform all of the catalogs into the comoving frame of WLM. This is the frame in which WLM's stars are---on average---stationary between epochs.
HubPUG first rotates the distortion-corrected positions of each exposure so that its axes point east and north (using the orientations in the \texttt{fits} image headers) and then solves for the 6DLT from each exposure onto the frame of the first first-epoch exposure, using the template-fit positions of the frame stars. Since every catalog is built from the same lists of stars and galaxies, sources are matched between exposures by their identifiers. The frame stars are refined iteratively, and a star is used only if it is measured in at least three quarters of all exposures. In each iterative round, any star whose mean offset from its position in the reference exposure or whose scatter between exposures exceeds a set tolerance is dropped before the transformations are solved again. The tolerance starts at 5~pixels and shrinks by a factor of 1.25 in each round, down to 0.05~pixels (2.5~mas, or 0.28--0.36~mas~yr$^{-1}$ over our baselines), so that the final transformations are set by the stars whose positions are most consistent with the bulk motion of WLM. We apply the final transformations to every star and galaxy.

For the full-field HST-to-HST measurement, this leaves 319 transformation frame stars (308--319 in each exposure), with rms residuals per coordinate (between a frame star's transformed position and its position in the reference exposure) of 0.028--0.032~pixels (1.4--1.6~mas) for the first-epoch exposures and 0.035--0.039~pixels (1.8--2.0~mas) for the second-epoch exposures. For each NIRCam unit, stars measured only in the first-epoch exposures also meet the 75\% requirement, since the 26 first-epoch exposures make up 87\% of the 30. The first-epoch transformations are therefore set by 399--423 stars from across the ACS field (with rms residuals of 0.029--0.034~pixels), while each NIRCam exposure's transformation is set by the 26--75 of these stars that fall on its detector, with rms residuals of 0.016--0.025~pixels (0.8--1.3~mas). These residuals are, as expected, larger than those of the registrations in Sections~\ref{ssec:frame} and \ref{ssec:register}, which use PSF-fit positions of bright stars: the template-fit positions are less precise, the frame stars extend to ${\rm F814W} = 26$, and each residual is measured against the star's position in a single reference exposure.

\subsection{Per-galaxy Proper Motions and Their Uncertainties} \label{ssec:galpm}

In the comoving frame, we take the position of each galaxy in each epoch to be the mean of its positions in that epoch's exposures. We require that the galaxy be measured in at least 75\% of those exposures (three of the four exposures of a NIRCam unit, twenty of the twenty-six first-epoch exposures, or five of the six second-epoch ACS exposures), and that the uncertainty of its mean position (see below) be less than 1~pixel.
The difference between a galaxy's positions in the two epochs, divided by the time baseline (7.02~years for the F090W and F150W NIRCam exposures, 9.02~years for the F200W NIRCam exposures, and 8.93~years for the second-epoch ACS exposures), is the galaxy's reflex motion, and its negative is an independent measurement of WLM's systemic PM. In what follows, we refer to this measurement as the galaxy's PM. Each galaxy's PM is then corrected for the residual motions of WLM's stars in the same part of the field (Section~\ref{ssec:localcorr}).

\begin{figure*}
\centering
\includegraphics[width=\textwidth]{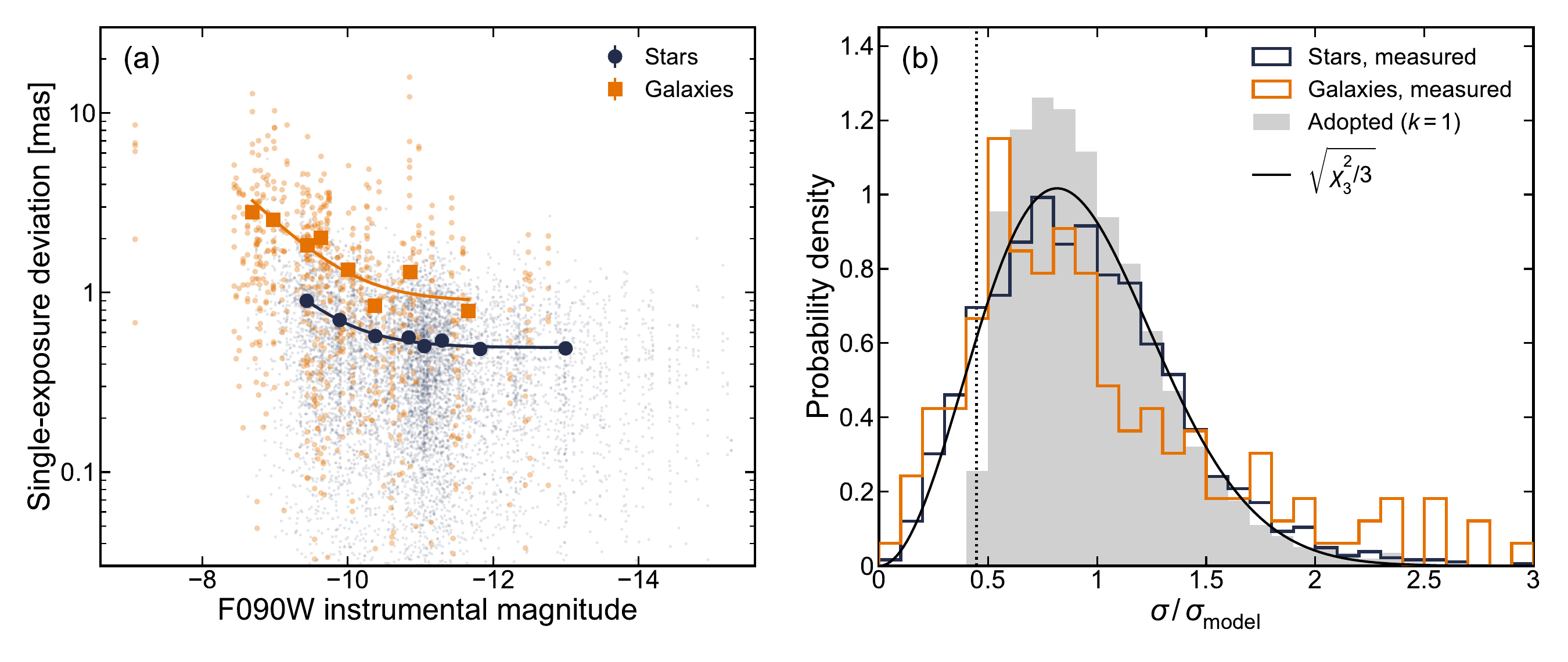}
\caption{The error model described in Section~\ref{ssec:galpm}, for the second epoch of the B2 F090W detector (four exposures). (a) Deviations of each source's single-exposure positions from its mean position (both coordinates, scaled by $\sqrt{n/(n-1)}$), against instrumental magnitude, for stars (small blue points) and galaxies (small orange points). The large symbols are the robust scatter in bins of magnitude, and the curves are the fitted $\sigma_1(m)$. (b) The ratio of each four-exposure source's measured uncertainty to its model uncertainty, for stars and galaxies, and of the adopted ($k = 1$) uncertainty to the model uncertainty (gray). The black curve is the distribution expected from sampling noise alone for four Gaussian positions, $\sqrt{\chi^2_3/3}$, and the dotted line marks the floor of the adopted uncertainties, $1/\sqrt{5}$ of the model uncertainty.}
\label{fig:errmodel}
\end{figure*}

Because each NIRCam unit has only four exposures, and because \citet{Warfield2026} showed (using $\chi^2$) that the per-galaxy PM uncertainties may be underestimated, we have added an error model to our position uncertainties.
For each source in each epoch, we start from the standard error of the mean of its $n$ positions, $\sigma_{\rm measured} = s/\sqrt{n}$, where $s$ is their sample standard deviation. For $n = 4$, this estimate is itself noisy (the sample variance of four values scatters by $\sim$80\% about the true variance), and sources whose positions happen to agree closely are assigned uncertainties much smaller than their true ones, which then dominate any inverse-variance-weighted mean.
We therefore combine each source's measured uncertainty with a model of the single-exposure scatter as a function of magnitude, which we fit separately to the galaxies and to the stars in each epoch of each run.
For every source measured in at least two exposures, we take the deviations of its positions from their mean, scaled by $\sqrt{n/(n-1)}$ so that their variance is the single-exposure variance, and fit their robust scatter in bins of magnitude (Figure~\ref{fig:errmodel}a) with
\begin{equation}
    \sigma_1(m)^2 = a^2 + b^2\,10^{0.8\,(m - m_{\rm ref})},
\end{equation}
a floor plus a term that is inversely proportional to the source's flux, as expected for faint sources whose noise is dominated by the sky. The model uncertainty of a source's mean position is then $\sigma_{\rm model} = \sigma_1(m)/\sqrt{n}$, and the uncertainty we adopt is
\begin{equation}
\sigma^2 = \frac{n\,\sigma_{\rm measured}^2 + k\,\sigma_{\rm model}^2}{n+k},
\end{equation}
the average of the two variances, weighted as if the model were $k$ additional exposures. For four exposures, the spread of $\sigma_{\rm measured}$ about the model is consistent with sampling noise alone (Figure~\ref{fig:errmodel}b). For F090W B2, $\sigma_{\rm measured}/\sigma_{\rm model}$ has a median of 0.89 for stars and 0.86 for galaxies, against 0.89 expected for four Gaussian positions, and 16\% of stars and 18\% of galaxies fall below half of the model uncertainty against 14\% expected.
We adopt $k = 1$, for which the model acts as a soft floor: however closely a source's own positions happen to agree, its uncertainty cannot fall below $\sigma_{\rm model}/\sqrt{n+1}$. For the 26 first-epoch exposures, each source's measured scatter dominates.
The uncertainty of each galaxy's PM is then the quadrature sum of the uncertainties of its two positions and of its local correction (Section~\ref{ssec:localcorr}), divided by the baseline, and we use these uncertainties both to cut and to weight the galaxies when combining them (Section~\ref{ssec:combine}).

\subsection{Local Correction} \label{ssec:localcorr}

As in Section~4 of \citet{Warfield2026}, we correct each galaxy's PM for the residual motion of WLM's stars in the same part of the field, which absorbs time-dependent offsets between the distortion-corrected frames of the two epochs, including those from imperfect CTE correction in the ACS images. The correction for a galaxy is the weighted, 3$\sigma$-clipped mean residual motion of the local-correction stars (Section~\ref{ssec:catalogs}) in the same first-epoch ACS quadrant, after dropping any star whose motion differs from the median of these stars by $\geq 2$~mas~yr$^{-1}$, and its uncertainty is added in quadrature to that of the galaxy (Section~\ref{ssec:galpm}). A star is used for the local correction only if it has a template-fit position in the reference exposure, in at least 22 of the 26 first-epoch exposures, and in at least four of the six second-epoch ACS exposures or three of the four exposures of a NIRCam unit. In the full-field HST-to-HST measurement, 5631 of the 14,531 selected stars meet this requirement (Figure~\ref{fig:cmds}). They are all brighter than ${\rm F814W} = 25.9$, with a median of 24.3. Each NIRCam unit has 302--878 such stars, and 97\% of those of the F090W and F200W units are also local-correction stars in the HST-to-HST measurements of the same areas. Unlike Draco~II, which has only $\sim 200$ such stars, WLM has thousands, and so we can also require them to be similar in brightness to the galaxy. We use only stars within 1~mag of the galaxy's second-epoch template-fit magnitude, and require at least four of them. This matches each galaxy to stars with similar CTE losses, as CTE---as well as other potential systematics associated with NIRCam's IR detectors that we do not test separately, but which may also be accounted for by the local correction---scales as a function of flux.

\subsection{Galaxy Selection and Combination} \label{ssec:combine}

Before combining the galaxies of a run, we apply five cuts. First, as in \citet{Warfield2026}, we drop galaxies whose total PM uncertainty is 1~mas~yr$^{-1}$ or more.
Second, we drop galaxies whose total PM magnitude exceeds 1~mas~yr$^{-1}$ at more than 3$\sigma$ (at WLM's distance, this corresponds to an implausible tangential velocity of $\sim$4700~km~s$^{-1}$).
Third, we keep only galaxies with a template quality of fit (Section~\ref{ssec:fits}) of $q < 0.25$ in the first epoch.
Fourth, we keep only galaxies with a stellar fraction of $f_\star < 0.5$ (Section~\ref{ssec:crowding}).
Fifth, we iteratively remove galaxies whose PMs differ from the weighted mean by more than $3\sigma$. The same cuts are applied to every run, but because the uncertainties differ between second epochs, the galaxies that pass can differ; the per-galaxy comparison of Section~\ref{ssec:res_overview} uses only the galaxies common to both.

We combine the remaining galaxies with a hierarchical weighted mean. Galaxies that share a local correction are first averaged with weights from their own uncertainties, the uncertainty of their common correction is then added once, and the groups are combined. Because each galaxy's local-correction stars are chosen by its own magnitude, in practice every galaxy is its own group, and this reduces to an inverse-variance-weighted mean. When $\chi^2_\nu > 1$, we scale the uncertainty of the mean by $\sqrt{\chi^2_\nu}$. The error model of Section~\ref{ssec:galpm} concerns the uncertainty of a position within an epoch, whereas this scaling accounts for any scatter between the epochs beyond those uncertainties. For each NIRCam filter, we combine the units into a single measurement by concatenating their galaxies, each corrected for its unit's injection bias (Section~\ref{ssec:injection}), and repeating the clipping and the mean on the combined table. For comparison, we also measure the HST-to-HST PM over the area of each NIRCam unit, using the galaxies and stars that are measured in at least two of that unit's NIRCam exposures, and combine these areas in the same way. Finally, we divide the SW and HST-to-HST PMs by the scale correction of Section~\ref{ssec:injection} and propagate its uncertainty; we do not apply this correction to the LW PMs or to the HST-to-HST PMs that we compare them with.

\subsection{Injection Tests and Bias Corrections} \label{ssec:injection}

To measure the biases of the HST-to-JWST measurements, we injected synthetic, at-rest galaxies into the real exposures and passed them through the full pipeline. For each NIRCam unit, in each of five realizations, we injected about 150 galaxies (defined by typical S\'ersic profiles) and 60 point sources into the first-epoch ACS exposures and the corresponding NIRCam exposures, with the injected point sources as the only stars used in the analysis. Since each injected galaxy is at rest relative to the injected point sources, the PM measured in each run is its bias. The injected galaxies were placed at least $0.5''$ from every \texttt{hst1pass} detection brighter than the median, and so the cut on $f_\star$ (Section~\ref{ssec:crowding}) removes only about 1\% of galaxies placed in this way.
Pooled over all NIRCam detectors of each filter, the total measured injection biases are $\leq$0.02~mas~yr$^{-1}$ (F090W: $\Delta\mu_{\alpha}^* = +0.010$~mas~yr$^{-1}$, $\Delta\mu_{\delta} = -0.018$~mas~yr$^{-1}$; F150W: $-0.005$, $-0.012$; F200W: $+0.016$, $-0.007$), while those of individual detectors reach $\sim$0.05~mas~yr$^{-1}$ (e.g., for B2 F090W $\Delta\mu_\delta = -0.044$~mas~yr$^{-1}$). We have not identified the origin of these biases. One possibility is that the different sampling of extended sources by the ACS and NIRCam pixel grids biases their fitted centers.

Because these per-unit biases are reproducible between realizations and show no trend with galaxy size, we subtract each unit's bias from its HST-to-JWST PMs, and add its uncertainty (the larger of the scatter between realizations and their mean quoted uncertainty, divided by $\sqrt{5}$) in quadrature. We do not correct the HST-to-HST measurements.
For the LW filters, the measured injection biases came out consistent with zero within uncertainties, but with uncertainties twice as large as the runs' own errors. Therefore, no bias is corrected for in the LW measurements.

However, galaxies injected at rest cannot reveal a pull toward zero motion, and so cannot be used to test for the potential effects discussed in Section~\ref{ssec:crowding}.
We therefore also injected galaxies that move by known amounts ($|\mu| \approx 1.35$ and 0.11~mas~yr$^{-1}$) into the E2 and E3 NIRCam exposures and the second-epoch ACS exposures. We ran six realizations of this test for each group of detectors, placing the galaxies away from bright stars (as above) in half of them and without regard to the stars in the other half, and fit the measured PMs as $k \times \mu_{\rm true} + b$. With the cut on $f_\star$ applied, the motions of the galaxies placed away from bright stars are recovered at full scale ($k = 0.98$--$1.01$), and those of the galaxies placed without regard to the stars at $k = 0.95$--$0.98$ for F090W, F150W, and HST-to-HST ($0.99$--$1.01$ for F200W). For the latter, the F090W $\mu_\delta$ offset reaches $b = -0.052 \pm 0.012$~mas~yr$^{-1}$. Over all placements, the HST-to-HST measurement has $k = 0.977 \pm 0.006$ and $0.992 \pm 0.009$, and an offset of $b = +0.008 \pm 0.004$ and $+0.019 \pm 0.007$~mas~yr$^{-1}$ in $\mu_\alpha^*$ and $\mu_\delta$. We do not apply this offset, but mention this as a possible systematic uncertainty of the HST-to-HST measurement.

However, these scales overstate how well a motion as small as WLM's is recovered in a crowded stamp. They are set mostly by the realizations with $|\mu| \approx 1.35$~mas~yr$^{-1}$, in which a galaxy whose motion has been pulled toward zero stands apart from the others and is removed by our 3$\sigma$ clip. For HST-to-HST, the clip removes 64\% of the galaxies with $f_\star \geq 0.1$ that were placed without regard to the stars in these realizations, and none of them in the 0.11~mas~yr$^{-1}$ realizations. Before the clip, the fraction of the injected motion that is recovered falls steeply with $f_\star$, as the gradient-power argument of Section~\ref{ssec:crowding} predicts: for HST-to-HST, from 0.99 for $f_\star < 0.02$ to 0.65 for $0.05 \leq f_\star < 0.1$, 0.34 for $0.1 \leq f_\star < 0.2$, and $\leq$0.06 for $f_\star \geq 0.2$.

We therefore correct each of our SW PMs for this loss by dividing it by the fraction of the motion that we expect it to recover, $s$. We measure $s$ from the injected galaxies that were placed without regard to the stars, because the background galaxies are unrelated to WLM's stars and fall among them at random, as these injected galaxies do. For each PM, $s$ is the fraction of the injected motion that these galaxies recover before the clip, using only the injected galaxies on the same detector and in the same filter that pass our cut on $f_\star$. The injection bias, where we apply one, is subtracted before the PM is divided by $s$.

One difference between the injected and the real galaxies has to be allowed for. For an injected galaxy, $f_\star$ counts only starlight, while for a real galaxy it also counts parts of the galaxy itself (Section~\ref{ssec:crowding}). The same cut therefore removes real galaxies that have less starlight than the injected galaxies it removes, and the $s$ above is slightly too low. At the other extreme, in which all of the $f_\star$ of the real galaxies is their own light, $s$ is higher by 0.02--0.07. We adopt the mean of the two values, and include half of their difference in the uncertainty of $s$.

For the full-field HST-to-HST measurement, this gives $s = 0.86 \pm 0.02$. Here we also weight the injected galaxies to match the stellar density around the real ones, because the two samples cover different parts of the field. The individual detectors that we use have $s$ between 0.77 and 0.96 (Table~\ref{tab:results}), and E2 B4 has $s = 0.70$--$0.74$. For a combined measurement, $s$ is the mean of its detectors' values, weighted as their galaxies are in the combined PM. The uncertainty of $s$ is included in the uncertainties of the corrected PMs. We do not correct the LW measurements, for which we have no moving-galaxy test.

The real galaxies give a check of this correction that does not depend on the injected ones. In the full-field HST-to-HST measurement, the 145 galaxies with $f_\star < 0.5$ move at $1.03^{+0.14}_{-0.13}$ times the PM of the 60 galaxies with $f_\star < 0.05$, where the injected galaxies predict 0.89. The two agree at about the 1$\sigma$ level, but the real galaxies alone cannot distinguish this loss from none. The correction also assumes that the fractional loss is the same for a motion of 0.1~mas~yr$^{-1}$ as for one of 1.35~mas~yr$^{-1}$ (the 0.11~mas~yr$^{-1}$ realizations alone give $0.90 \pm 0.07$ for HST-to-HST without the cut, against $0.77 \pm 0.02$ for the others), and that a star pulls an injected S\'ersic profile as it does a real galaxy (at the same magnitude, the positions of the real galaxies are 5--20\% more precise). Both would make the true loss smaller than the one that we correct for, and so we also list the PMs without the scale correction in Table~\ref{tab:results}.

\section{Results} \label{sec:results}

Table~\ref{tab:results} lists the PMs of WLM that we measure in each subfield against each second epoch, first as measured, then with the injection-bias correction of Section~\ref{ssec:injection}, and then with its scale correction as well. Unless we say otherwise, the SW PMs that we quote below include both corrections. We first present the systemic PM of WLM from the two ACS epochs (Section~\ref{ssec:systemic}), which is our most precise measurement and the reference for the HST-to-JWST PMs. We then present the HST-to-JWST PMs from the SW (Section~\ref{ssec:res_overview}) and LW (Section~\ref{ssec:res_lw}) detectors, and compare the precision of the second epochs (Section~\ref{ssec:precision}).

\subsection{The Systemic PM of WLM} \label{ssec:systemic}

Our most precise measurement is the HST-to-HST PM of the full ACS field, $(\mu_\alpha^*, \mu_\delta) = (+0.087~\pm~0.019,\allowbreak -0.079~\pm~0.017)$~mas~yr$^{-1}$ from 145 galaxies over 8.93~years,\footnote{$(+0.075~\pm~0.016,\allowbreak -0.068~\pm~0.014)$~mas~yr$^{-1}$ before the scale correction.} which we adopt as the systemic PM of WLM. At the distance of WLM, $989.5 \pm 26.9$~kpc ($\mu_0 = 24.977 \pm 0.059$; \citealt{Yan2025}, as listed in the Local Volume Database, \citealt{Pace2025}), a PM of 0.01~mas~yr$^{-1}$ corresponds to 47~km~s$^{-1}$. Our adopted PM then corresponds with a heliocentric tangential velocity of $(V_\alpha, V_\delta)_\odot = (+409 \pm 88,\allowbreak -372 \pm 79)$~km~s$^{-1}$. After subtracting the reflex of the solar motion, for which we adopt a solar velocity of $(12.9, 245.6, 7.8)$~km~s$^{-1}$ \citep{Drimmel2018}, this is $(V_\alpha, V_\delta)_{\rm MW} = (+287 \pm 88,\allowbreak -167 \pm 79)$~km~s$^{-1}$.

\begin{figure}
\centering
\includegraphics[width=\columnwidth]{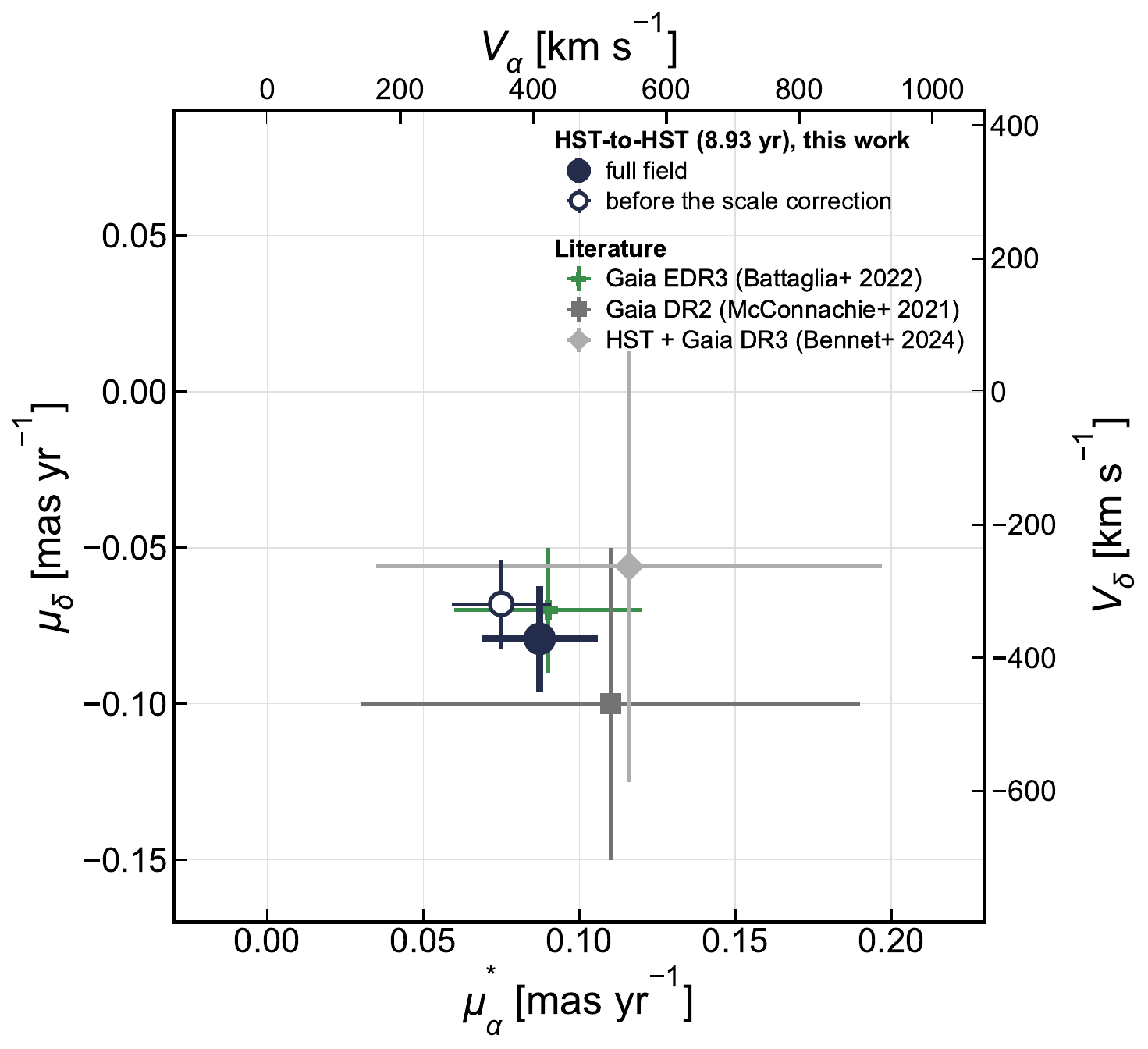}
\caption{Systemic PM of WLM from the two ACS epochs over the full ACS field (filled circle, with both corrections; open circle, before the scale correction), compared with the Gaia EDR3 PM of \protect\cite{Battaglia2022}, the Gaia DR2 PM of \protect\cite{McConnachie2021}, and the HST and Gaia DR3 PM of \protect\cite{Bennet2024}. The top and right axes give the heliocentric tangential velocity at the adopted distance of WLM.}
\label{fig:pm_systemic}
\end{figure}

In Figure~\ref{fig:pm_systemic}, we compare this PM with previous measurements. It agrees with the Gaia EDR3 PM of \cite{Battaglia2022}, $(+0.090~\pm~0.030,\allowbreak -0.070~\pm~0.020)$~mas~yr$^{-1}$, to within 0.4$\sigma$ in each component, with an uncertainty that is about 40\% smaller in $\mu_\alpha^*$ and 15\% smaller in $\mu_\delta$ than their statistical uncertainties (\citealt{Battaglia2022} also list a systematic uncertainty of 0.021~mas~yr$^{-1}$ per component for WLM). It also agrees to within 0.4$\sigma$ with the Gaia DR2 PM of \cite{McConnachie2021}, $(+0.11~\pm~0.08,\allowbreak -0.10~\pm~0.05)$~mas~yr$^{-1}$, and with the PM of \citet{Bennet2024}, $(+0.116~\pm~0.081,\allowbreak -0.056~\pm~0.069)$~mas~yr$^{-1}$, from a combination of HST imaging with the Gaia DR3 catalog. The uncertainties of both are three to four times larger than ours. Our PM is therefore consistent with all three and is the most precise of them. It is also the only one that does not rely on Gaia, as its reference frame is set by background galaxies.

\subsection{HST-to-JWST PMs from the SW Detectors} \label{ssec:res_overview}

\begin{figure}
\centering
\includegraphics[width=\columnwidth]{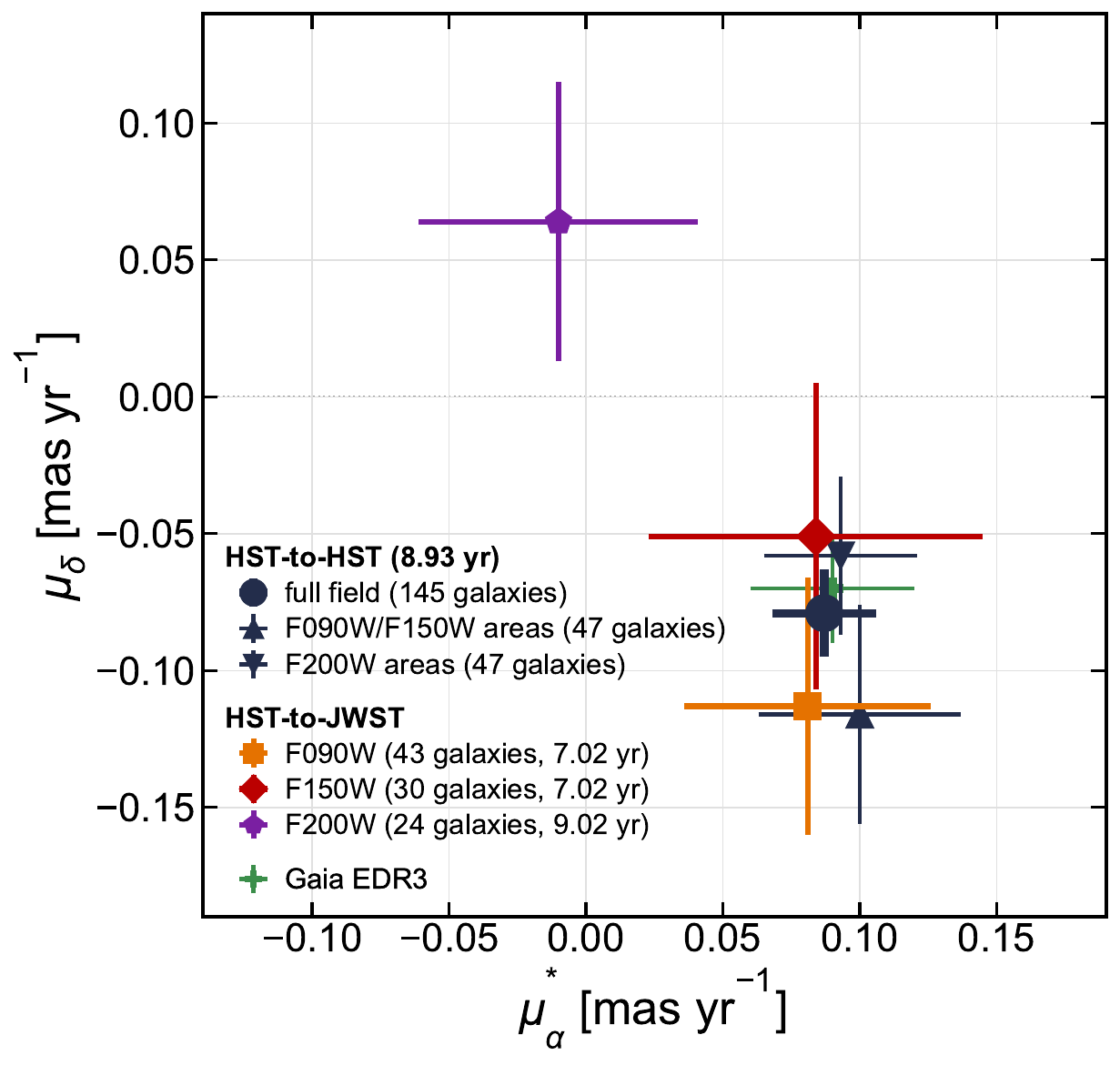}
\caption{PMs of WLM from HST, Gaia, and the NIRCam SW detectors. We show the HST-to-HST PM of the full ACS field, the HST-to-HST PMs over the areas of the SW detectors, and the injection-corrected HST-to-JWST PMs from the combined tables of each SW filter (E2 B1--B3 for F090W and F150W; E3 A3, A4, B3, and B4 for F200W), compared with the Gaia EDR3 PM of \protect\cite{Battaglia2022}. The number of galaxies and the time baseline of each measurement are given in the legend. All of our PMs include the scale correction of Section~\ref{ssec:injection}.}
\label{fig:pm_sw}
\end{figure}

\begin{figure*}
\centering
\includegraphics[width=\textwidth]{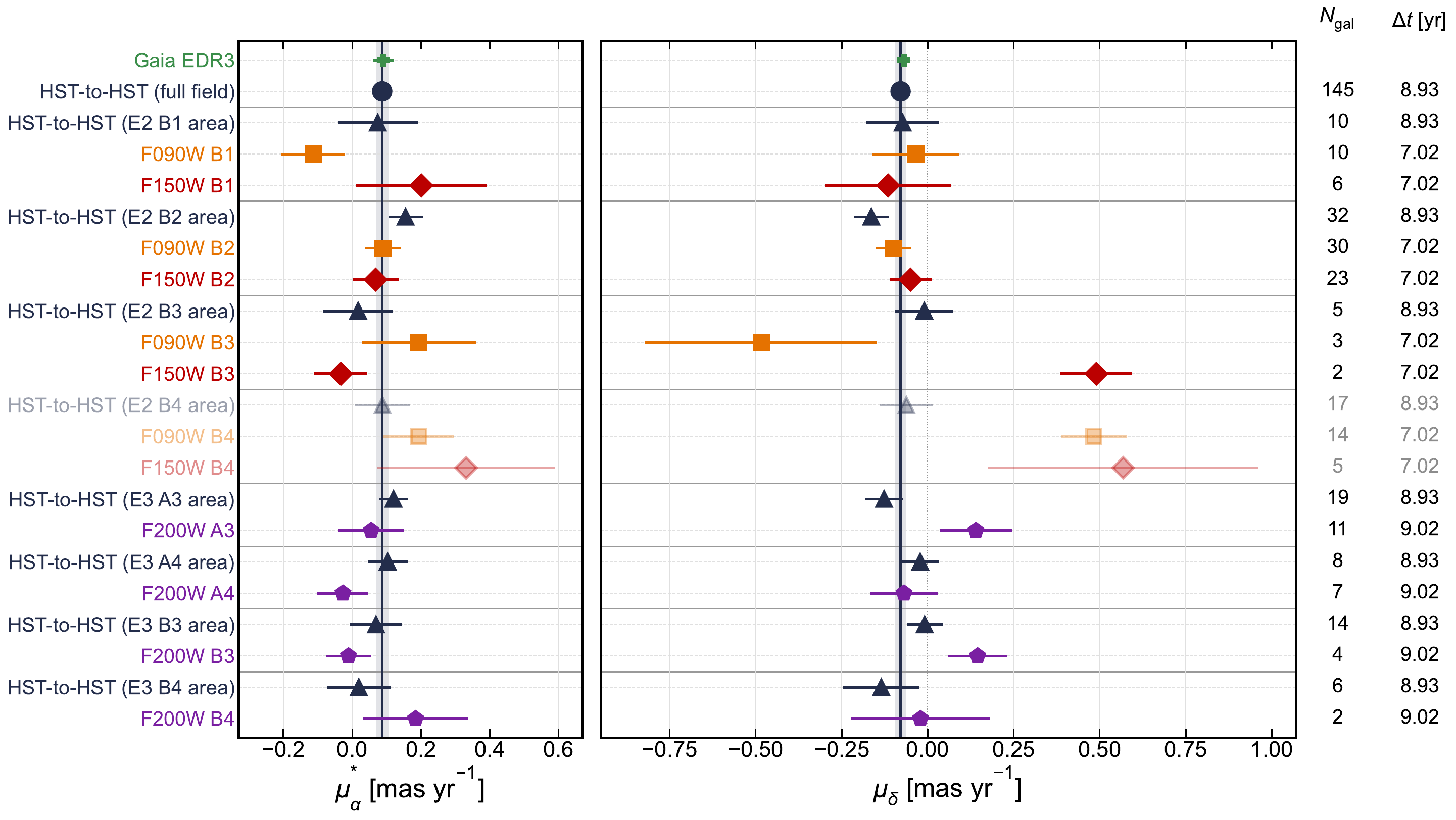}
\caption{Recovered values for $\mu_\alpha^*$ (left) and $\mu_\delta$ (right) of WLM for each NIRCam SW detector and filter separately, with each HST-to-JWST PM below the HST-to-HST PM for the equivalent area. The top two rows are the Gaia EDR3 PM of \protect\cite{Battaglia2022} and the full-field HST-to-HST PM; the latter is repeated in both panels as the vertical line, with its uncertainty as the shaded band. The two panels have the same scale. The columns on the right give the number of galaxies and the time baseline of each measurement. The E2 B4 rows (faded), which we do not use in the combined results, are shown for comparison. All of our PMs include the scale correction of Section~\ref{ssec:injection}.}
\label{fig:pm_sw_panels}
\end{figure*}

Figures~\ref{fig:pm_sw} and \ref{fig:pm_sw_panels} compare the SW measurements with each other, with the HST-to-HST PMs over the same areas, and with the Gaia-based PM of \citet{Battaglia2022}.

\begin{deluxetable*}{llcccccccccc}
\tablewidth{0pt}
\tabletypesize{\fontsize{5.6pt}{6.8pt}\selectfont}
\setlength{\tabcolsep}{1.6pt}
\tablecaption{Proper Motions of WLM by Subfield and Second Epoch\label{tab:results}}
\tablehead{
\colhead{} & \colhead{} & \colhead{} & \multicolumn{3}{c}{Measured} & \multicolumn{3}{c}{With injection-bias correction} & \multicolumn{3}{c}{With both corrections} \\
\cline{4-6} \cline{7-9} \cline{10-12}
\colhead{Subfield} & \colhead{Second Epoch} & \colhead{$\Delta t$} & \colhead{$N_{\rm gal}$} & \colhead{$\mu_\alpha^*$} & \colhead{$\mu_\delta$} & \colhead{$N_{\rm gal}$} & \colhead{$\mu_\alpha^*$} & \colhead{$\mu_\delta$} & \colhead{$s$} & \colhead{$\mu_\alpha^*$} & \colhead{$\mu_\delta$} \\
\colhead{} & \colhead{} & \colhead{(yr)} & \colhead{} & \colhead{(mas~yr$^{-1}$)} & \colhead{(mas~yr$^{-1}$)} & \colhead{} & \colhead{(mas~yr$^{-1}$)} & \colhead{(mas~yr$^{-1}$)} & \colhead{} & \colhead{(mas~yr$^{-1}$)} & \colhead{(mas~yr$^{-1}$)}
}
\startdata
E2 B1 & NIRCam F090W & 7.02 & 10 & $-0.097 \pm 0.084$ & $-0.035 \pm 0.115$ & 10 & $-0.105 \pm 0.085$ & $-0.032 \pm 0.116$ & $0.92 \pm 0.04$ & $-0.114 \pm 0.092$ & $-0.035 \pm 0.126$ \\
 & NIRCam F150W & 7.02 & 6 & $+0.147 \pm 0.170$ & $-0.073 \pm 0.164$ & 6 & $+0.180 \pm 0.170$ & $-0.103 \pm 0.165$ & $0.90 \pm 0.05$ & $+0.201 \pm 0.190$ & $-0.115 \pm 0.184$ \\
 & ACS F814W & 8.93 & 10 & $+0.061 \pm 0.096$ & $-0.060 \pm 0.087$ & \nodata & \nodata & \nodata & $0.83 \pm 0.04$ & $+0.074 \pm 0.116$ & $-0.073 \pm 0.105$ \\
\hline
E2 B2 & NIRCam F090W & 7.02 & 30 & $+0.095 \pm 0.042$ & $-0.129 \pm 0.037$ & 30 & $+0.077 \pm 0.044$ & $-0.085 \pm 0.043$ & $0.86 \pm 0.05$ & $+0.090 \pm 0.052$ & $-0.099 \pm 0.050$ \\
 & NIRCam F150W & 7.02 & 23 & $+0.055 \pm 0.056$ & $-0.080 \pm 0.050$ & 23 & $+0.059 \pm 0.058$ & $-0.043 \pm 0.053$ & $0.87 \pm 0.05$ & $+0.068 \pm 0.067$ & $-0.050 \pm 0.061$ \\
 & ACS F814W & 8.93 & 32 & $+0.128 \pm 0.041$ & $-0.135 \pm 0.041$ & \nodata & \nodata & \nodata & $0.82 \pm 0.04$ & $+0.155 \pm 0.050$ & $-0.164 \pm 0.051$ \\
\hline
E2 B3 & NIRCam F090W & 7.02 & 3 & $+0.174 \pm 0.134$ & $-0.440 \pm 0.274$ & 3 & $+0.160 \pm 0.135$ & $-0.399 \pm 0.275$ & $0.82 \pm 0.07$ & $+0.194 \pm 0.165$ & $-0.484 \pm 0.336$ \\
 & NIRCam F150W & 7.02 & 2 & $-0.009 \pm 0.059$ & $+0.348 \pm 0.070$ & 2 & $-0.026 \pm 0.061$ & $+0.387 \pm 0.074$ & $0.79 \pm 0.07$ & $-0.033 \pm 0.077$ & $+0.490 \pm 0.104$ \\
 & ACS F814W & 8.93 & 5 & $+0.013 \pm 0.078$ & $-0.008 \pm 0.065$ & \nodata & \nodata & \nodata & $0.77 \pm 0.06$ & $+0.017 \pm 0.101$ & $-0.010 \pm 0.085$ \\
\hline
E2 SW combined & NIRCam F090W & 7.02 & 43 & $+0.086 \pm 0.037$ & $-0.136 \pm 0.035$ & 43 & $+0.070 \pm 0.039$ & $-0.098 \pm 0.040$ & $0.86 \pm 0.05$ & $+0.081 \pm 0.045$ & $-0.113 \pm 0.046$ \\
 & NIRCam F150W & 7.02 & 30 & $+0.066 \pm 0.052$ & $-0.075 \pm 0.046$ & 30 & $+0.073 \pm 0.053$ & $-0.044 \pm 0.049$ & $0.87 \pm 0.05$ & $+0.084 \pm 0.062$ & $-0.051 \pm 0.057$ \\
 & ACS F814W & 8.93 & 47 & $+0.081 \pm 0.030$ & $-0.094 \pm 0.032$ & \nodata & \nodata & \nodata & $0.81 \pm 0.03$ & $+0.100 \pm 0.037$ & $-0.116 \pm 0.040$ \\
\hline
E2 B5 & NIRCam F250M & 7.02 & 23 & $+0.120 \pm 0.090$ & $-0.136 \pm 0.083$ & \nodata & \nodata & \nodata & \nodata & \nodata & \nodata \\
 & NIRCam F430M & 7.02 & 21 & $+0.112 \pm 0.092$ & $-0.162 \pm 0.091$ & \nodata & \nodata & \nodata & \nodata & \nodata & \nodata \\
 & ACS F814W & 8.93 & 66 & $+0.059 \pm 0.026$ & $-0.073 \pm 0.026$ & \nodata & \nodata & \nodata & \nodata & \nodata & \nodata \\
\hline
E3 A3 & NIRCam F200W & 9.02 & 11 & $+0.028 \pm 0.084$ & $+0.122 \pm 0.090$ & 11 & $+0.049 \pm 0.085$ & $+0.125 \pm 0.095$ & $0.90 \pm 0.06$ & $+0.055 \pm 0.095$ & $+0.139 \pm 0.106$ \\
 & ACS F814W & 8.93 & 19 & $+0.106 \pm 0.036$ & $-0.112 \pm 0.048$ & \nodata & \nodata & \nodata & $0.88 \pm 0.04$ & $+0.121 \pm 0.042$ & $-0.127 \pm 0.055$ \\
\hline
E3 A4 & NIRCam F200W & 9.02 & 7 & $-0.007 \pm 0.070$ & $-0.067 \pm 0.094$ & 7 & $-0.026 \pm 0.071$ & $-0.066 \pm 0.095$ & $0.96 \pm 0.04$ & $-0.027 \pm 0.074$ & $-0.069 \pm 0.099$ \\
 & ACS F814W & 8.93 & 8 & $+0.094 \pm 0.053$ & $-0.020 \pm 0.050$ & \nodata & \nodata & \nodata & $0.91 \pm 0.04$ & $+0.104 \pm 0.059$ & $-0.022 \pm 0.055$ \\
\hline
E3 B3 & NIRCam F200W & 9.02 & 4 & $+0.020 \pm 0.059$ & $+0.123 \pm 0.077$ & 4 & $-0.010 \pm 0.060$ & $+0.132 \pm 0.077$ & $0.91 \pm 0.04$ & $-0.011 \pm 0.066$ & $+0.146 \pm 0.086$ \\
 & ACS F814W & 8.93 & 14 & $+0.064 \pm 0.070$ & $-0.008 \pm 0.048$ & \nodata & \nodata & \nodata & $0.92 \pm 0.03$ & $+0.069 \pm 0.075$ & $-0.008 \pm 0.052$ \\
\hline
E3 B4 & NIRCam F200W & 9.02 & 2 & $+0.190 \pm 0.140$ & $-0.028 \pm 0.185$ & 2 & $+0.168 \pm 0.140$ & $-0.019 \pm 0.185$ & $0.92 \pm 0.05$ & $+0.184 \pm 0.154$ & $-0.021 \pm 0.202$ \\
 & ACS F814W & 8.93 & 6 & $+0.017 \pm 0.084$ & $-0.122 \pm 0.100$ & \nodata & \nodata & \nodata & $0.90 \pm 0.03$ & $+0.019 \pm 0.093$ & $-0.135 \pm 0.110$ \\
\hline
E3 SW combined & NIRCam F200W & 9.02 & 25 & $-0.019 \pm 0.047$ & $+0.079 \pm 0.046$ & 24 & $-0.009 \pm 0.047$ & $+0.059 \pm 0.047$ & $0.92 \pm 0.03$ & $-0.009 \pm 0.051$ & $+0.064 \pm 0.051$ \\
 & ACS F814W & 8.93 & 47 & $+0.084 \pm 0.025$ & $-0.052 \pm 0.026$ & \nodata & \nodata & \nodata & $0.90 \pm 0.02$ & $+0.094 \pm 0.027$ & $-0.058 \pm 0.029$ \\
\hline
E3 A5 & NIRCam F360M & 9.02 & 13 & $+0.247 \pm 0.101$ & $-0.068 \pm 0.094$ & \nodata & \nodata & \nodata & \nodata & \nodata & \nodata \\
 & ACS F814W & 8.93 & 32 & $+0.067 \pm 0.030$ & $-0.055 \pm 0.031$ & \nodata & \nodata & \nodata & \nodata & \nodata & \nodata \\
\hline
E3 B5 & NIRCam F360M & 9.02 & 8 & $+0.201 \pm 0.142$ & $-0.154 \pm 0.077$ & \nodata & \nodata & \nodata & \nodata & \nodata & \nodata \\
 & ACS F814W & 8.93 & 27 & $+0.064 \pm 0.041$ & $-0.038 \pm 0.036$ & \nodata & \nodata & \nodata & \nodata & \nodata & \nodata \\
\hline
\textbf{Full ACS field} & \textbf{ACS F814W} & \textbf{8.93} & \textbf{145} & $\mathbf{+0.075 \pm 0.016}$ & $\mathbf{-0.068 \pm 0.014}$ & \nodata & \nodata & \nodata & $\mathbf{0.86 \pm 0.02}$ & $\mathbf{+0.087 \pm 0.019}$ & $\mathbf{-0.079 \pm 0.017}$ \\
\enddata
\tablecomments{All measurements use E1 ACS F814W as the first epoch and the forward (blurred-image) kernels of Section~\ref{ssec:kernel}. $\Delta t$ is the baseline between E1 and the second epoch (Table~\ref{tab:obs}). $N_{\rm gal}$ is the number of galaxies that pass the selection of Section~\ref{ssec:combine}. The SW combined rows combine the measurements of the E2 B1--B3 detectors (which lie within the area of E2 B5) and of the E3 A3, A4, B3, and B4 detectors (Section~\ref{ssec:combine}). Uncertainties of the measured PMs are the statistical uncertainties of the weighted means, scaled by $\sqrt{\chi^2_\nu}$ where $\chi^2_\nu > 1$; with the injection-bias correction, the uncertainty of the subtracted bias is added in quadrature. For the combined measurements, this correction is applied to each galaxy, and so the galaxy selection, and $N_{\rm gal}$, can differ between the first two column groups. The ACS and NIRCam LW measurements are not injection-corrected (Section~\ref{ssec:injection}), and so have no entries in the second column group. $s$ is the fraction of an injected motion that is recovered by galaxies placed without regard to WLM's stars (Section~\ref{ssec:injection}). The PMs with both corrections are the injection-corrected PMs (the measured PMs, for ACS) divided by $s$, with the uncertainty of $s$ included. The NIRCam LW measurements, and the ACS measurements over the LW footprints, are not scale-corrected.}
\end{deluxetable*}

If we calculate the PM obtained by combining all of the individual background galaxy results for the F090W E2 B1--B3 detectors, we obtain $(\mu_\alpha^*, \mu_\delta) = (+0.081~\pm~0.045,\allowbreak -0.113~\pm~0.046)$~mas~yr$^{-1}$ from 43 galaxies. This agrees with the HST-to-HST PM over the same area, $(+0.100~\pm~0.037,\allowbreak -0.116~\pm~0.040)$~mas~yr$^{-1}$ from 47 galaxies, to within 0.02~mas~yr$^{-1}$ in both components ($\leq$0.3$\sigma$). The combined F150W PM, $(+0.084~\pm~0.062,\allowbreak -0.051~\pm~0.057)$~mas~yr$^{-1}$ from 30 galaxies, differs from it by 0.02 and 0.07~mas~yr$^{-1}$ (0.2 and 0.9$\sigma$). Both combined PMs also differ from our adopted PM (Section~\ref{ssec:systemic}) by less than 0.04~mas~yr$^{-1}$ in each component. Without the injection-bias correction, the F090W and F150W PMs are $(+0.100~\pm~0.043,\allowbreak -0.158~\pm~0.042)$ and $(+0.075~\pm~0.060,\allowbreak -0.086~\pm~0.054)$~mas~yr$^{-1}$, respectively, so that the correction changes the combined F090W $\mu_\delta$ by about its uncertainty, but leaves both filters consistent with the HST-to-HST PM. In velocity, the combined F090W PM, with uncertainties of 210--220~km~s$^{-1}$ per component, is the most precise of our HST-to-JWST measurements.

In the E3 fields, the combined F200W PM over A3, A4, B3, and B4, $(-0.009~\pm~0.051,\allowbreak +0.064~\pm~0.051)$~mas~yr$^{-1}$ from 24 galaxies, is farther from the HST-to-HST PM over the same area, $(+0.094~\pm~0.027,\allowbreak -0.058~\pm~0.029)$~mas~yr$^{-1}$ from 47 galaxies, by $-0.10$ and $+0.12$~mas~yr$^{-1}$ (1.8 and 2.1$\sigma$, Figure~\ref{fig:pm_sw}).

Because the two measurements of each field share their first epoch, the uncertainties of these differences are smaller than the quadrature sums we have used above, so this comparison is lenient. A more direct test is the difference between the HST-to-JWST and HST-to-HST PMs of each galaxy that is used in both measurements, in which most of the first-epoch errors cancel. Its weighted mean over the galaxies common to both measurements (with both corrections applied) is $(-0.110~\pm~0.056,\allowbreak +0.120~\pm~0.065)$~mas~yr$^{-1}$ for F090W (37 galaxies on B1--B3), $(-0.052~\pm~0.065,\allowbreak +0.170~\pm~0.076)$~mas~yr$^{-1}$ for F150W (27 galaxies on B1--B3), and $(-0.133~\pm~0.069,\allowbreak +0.125~\pm~0.063)$~mas~yr$^{-1}$ for F200W (22 galaxies on A3, A4, B3, and B4). In all three filters, then, the HST-to-JWST $\mu_\delta$ of the same galaxies is larger by 0.12--0.17~mas~yr$^{-1}$ (1.8--2.2$\sigma$). The HST-to-JWST $\mu_\alpha^*$ of the same galaxies is also smaller in all three filters, by 0.05--0.13~mas~yr$^{-1}$ (0.8--2.0$\sigma$). In F200W, the $\mu_\delta$ difference comes almost entirely from A3, where it is $+0.43~\pm~0.15$~mas~yr$^{-1}$ from 10 galaxies. Without the injection-bias correction, the F090W and F150W differences in $\mu_\delta$ are smaller, $+0.078~\pm~0.066$ and $+0.134~\pm~0.075$~mas~yr$^{-1}$.

\begin{figure*}
\centering
\includegraphics[width=\textwidth]{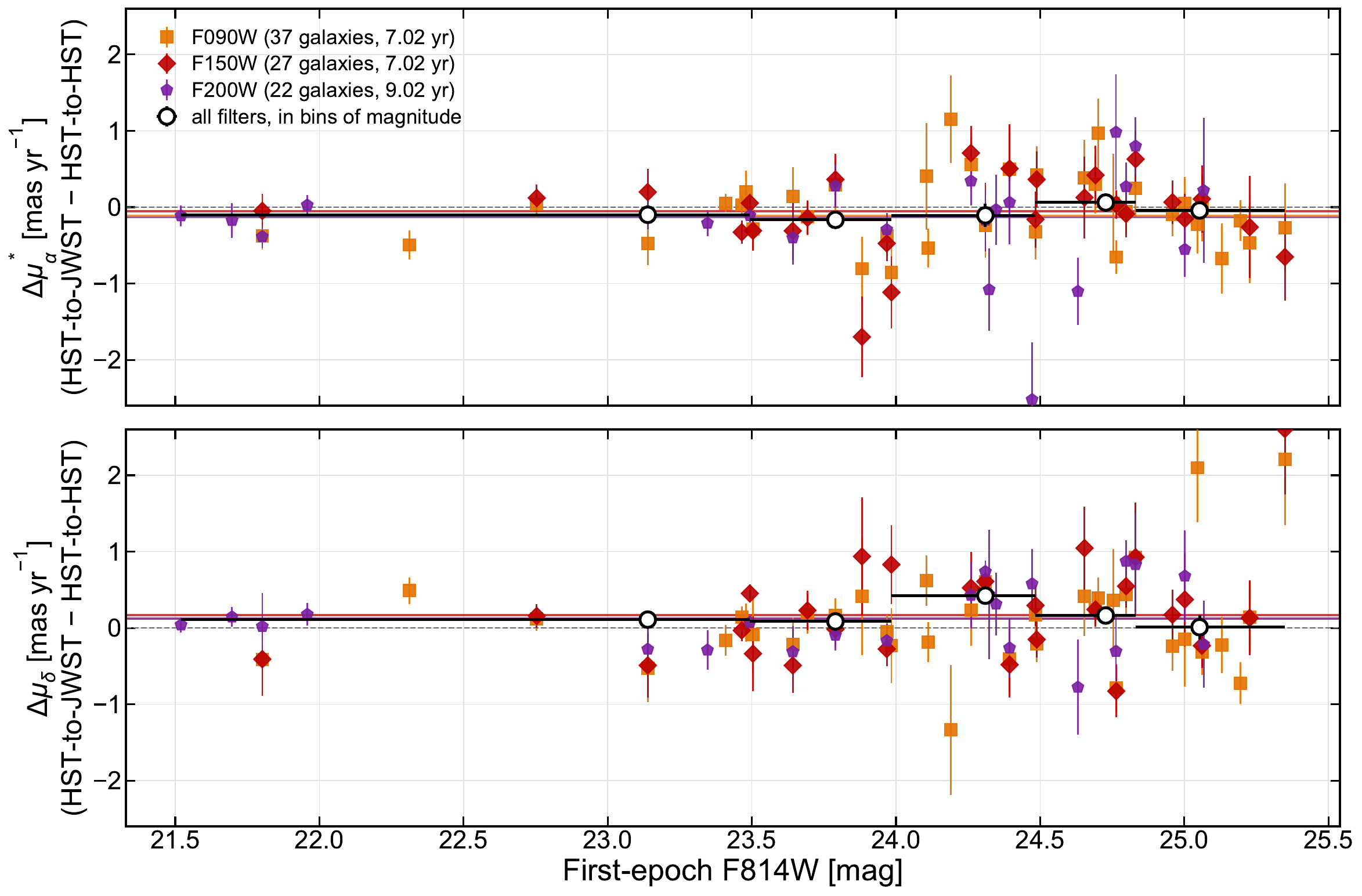}
\caption{Difference between the HST-to-JWST and HST-to-HST PMs of each galaxy used in both measurements over the same detector area, in $\mu_\alpha^*$ (top) and $\mu_\delta$ (bottom), against the galaxy's F814W magnitude in the first epoch. Colors and symbols mark the NIRCam filter (E2 B1--B3 for F090W and F150W; E3 A3, A4, B3, and B4 for F200W), with the number of galaxies and the time baseline in the legend; the HST-to-JWST PMs are injection-corrected, and both PMs are scale-corrected. Error bars are the uncertainties of the differences, with the covariance from the shared first-epoch positions removed. The colored horizontal lines are the weighted mean difference for each filter, and the open black circles are the weighted mean of all three filters in five bins of magnitude, whose widths are shown by the horizontal bars.}
\label{fig:galdiff_mag}
\end{figure*}

Figure~\ref{fig:galdiff_mag} shows these per-galaxy differences against the first-epoch F814W magnitude of each galaxy. There are no obvious trends with magnitude in either component. Pooling the three filters in five bins of magnitude, the mean difference stays between $-0.17$ and $+0.06$~mas~yr$^{-1}$ in $\mu_\alpha^*$ and between $+0.01$ and $+0.42$~mas~yr$^{-1}$ in $\mu_\delta$, with no monotonic change with magnitude, and the differences only scatter more for fainter galaxies, as their uncertainties do. Pooling the three filters, the differences also show no significant trend with the position of the galaxy on the ACS detector, and no dependence on the number of parallel transfers of the kind that residual CTE would produce (a slope of $+0.005 \pm 0.104$~mas~yr$^{-1}$ per 1000 transfers along the detector $y$-axis, which lies $14^\circ$ from the east--west direction).

Rather than any one property of the galaxies, three features of the E2 data limit these measurements. First, the number of usable galaxies is small. Of the $\sim 88$ and $\sim 65$ galaxies measured in each B2 and B1 exposure, 30 and 10 pass our selection in F090W, against 145 of 310 for the full ACS field, in part because of the crowding of these central fields. Second, because we process each NIRCam detector separately, each measurement is anchored only by the stars on its own detector (26--75 frame stars per NIRCam exposure, against 319 for the full ACS field), and in the most crowded fields few of them are isolated. Systematics tied to the stars of individual detectors, like the E2 B4 offset, therefore do not average out over the field as they can for ACS. Third, the E2 baseline of 7.02~years is shorter than the 8.93 and 9.02~years of E3, and much shorter than would be ideal for a galaxy at the distance of WLM ($\gtrsim 10$~years). Over 7.02~years, the PM of WLM ($|\mu| \approx 0.12$~mas~yr$^{-1}$) moves the background galaxies by only $\sim 0.8$~mas, or 0.017~ACS pixels (0.027~NIRCam SW pixels), which is on a comparable scale to the systematics we find above (the $+0.120$~mas~yr$^{-1}$ $\mu_\delta$ offset in F090W, for example, amounts to 0.8~mas over the same baseline). It is therefore reasonable to assume that these motions begin to test the limits of the data.

However, the individual detectors also scatter more than their uncertainties suggest (Figure~\ref{fig:pm_sw_panels}). In part, this is owed to counts: the E2 B3 measurements rest on only two or three galaxies, and the B1 measurements on six to ten.
The E2 B4 detector, which we do not use, shows the largest offsets: its F090W and F150W PMs have $\mu_\delta = +0.483~\pm~0.095$ and $+0.568~\pm~0.392$~mas~yr$^{-1}$, against $-0.062~\pm~0.077$~mas~yr$^{-1}$ for the HST-to-HST PM over the same area, and the per-galaxy differences give $+0.50~\pm~0.12$ and $+0.72~\pm~0.38$~mas~yr$^{-1}$ in $\mu_\delta$ (12 and 4 galaxies). We have not identified the cause of this offset. Because it is present in both filters, it is more likely tied to the detector or its part of the field than to wavelength.

\subsection{HST-to-JWST PMs from the LW Detectors} \label{ssec:res_lw}

\begin{figure}
\centering
\includegraphics[width=\columnwidth]{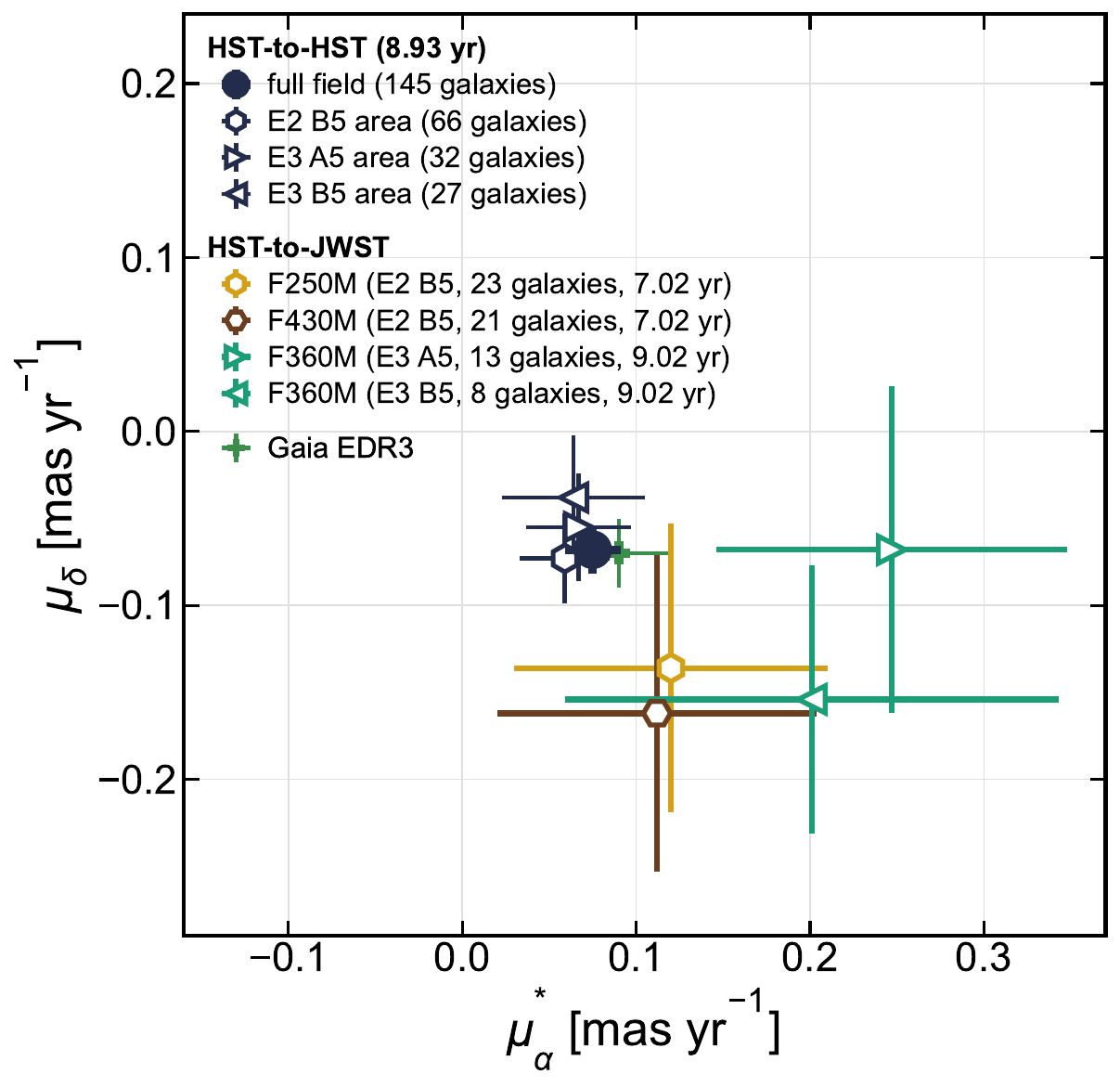}
\caption{PMs of WLM from the NIRCam LW detectors, with the HST and Gaia results for comparison. We show the HST-to-HST PM of the full ACS field and over each LW footprint, the HST-to-JWST PMs from the E2 B5 F250M and F430M exposures and from the E3 A5 and B5 F360M exposures, and the Gaia EDR3 PM of \protect\cite{Battaglia2022}. The number of galaxies and the time baseline of each measurement are given in the legend. None of our PMs in this figure is scale-corrected.}
\label{fig:pm_lw}
\end{figure}

\begin{figure*}
\centering
\includegraphics[width=\textwidth]{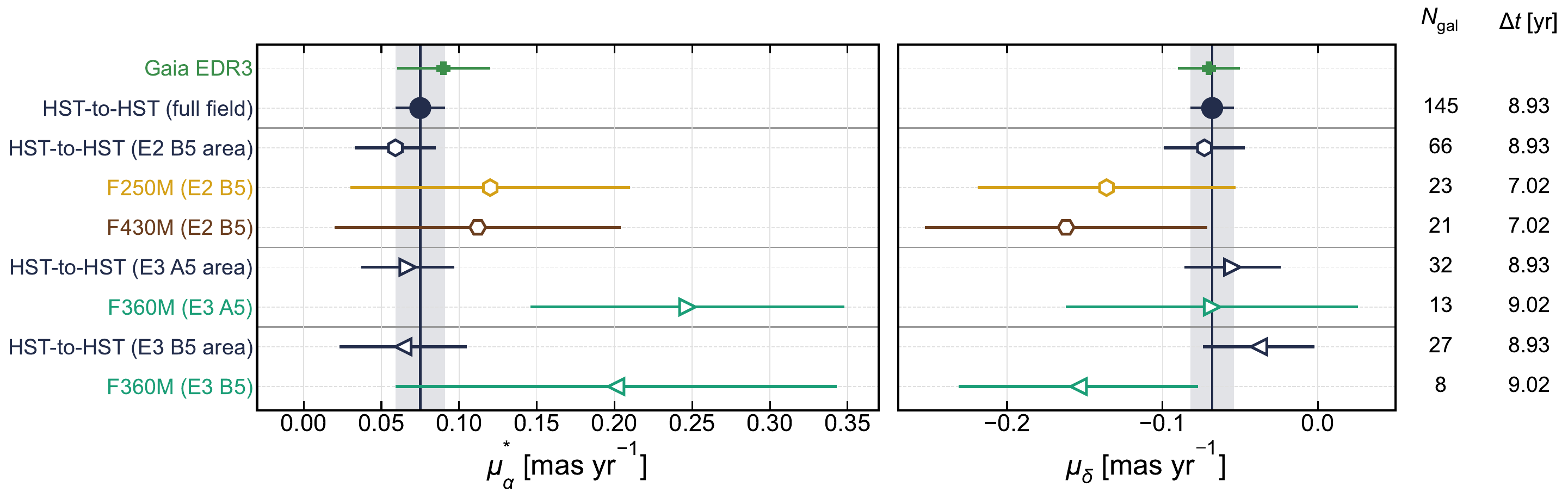}
\caption{Recovered values for $\mu_\alpha^*$ (left) and $\mu_\delta$ (right) of WLM for each NIRCam LW detector and filter separately, with each HST-to-JWST PM below the HST-to-HST PM for the equivalent area. As in Figure~\ref{fig:pm_sw_panels}, the top two rows are the Gaia EDR3 PM of \protect\cite{Battaglia2022} and the full-field HST-to-HST PM, the latter repeated in both panels as the vertical line and shaded band; the two panels have the same scale; and the columns on the right give the number of galaxies and the time baseline of each measurement. None of our PMs in this figure is scale-corrected.}
\label{fig:pm_lw_panels}
\end{figure*}

Figures~\ref{fig:pm_lw} and \ref{fig:pm_lw_panels} show the LW measurements in the same way.

The LW measurements are perhaps the most surprising and exciting. Because we expect large intrinsic changes in observed galaxy profiles across the F814W--LW color baselines, and because of the LW channel's coarser pixel scale, we predicted that images in these filters would not be useful for PM measurements.
However, though less precise, the LW results are consistent with the HST-to-HST PMs over the same footprints in the E2 field (Figure~\ref{fig:pm_lw}; none of the PMs in this comparison is scale-corrected): F250M gives $(+0.120~\pm~0.090,\allowbreak -0.136~\pm~0.083)$~mas~yr$^{-1}$ from 23 galaxies and F430M gives $(+0.112~\pm~0.092,\allowbreak -0.162~\pm~0.091)$~mas~yr$^{-1}$ from 21 galaxies, against $(+0.059~\pm~0.026,\allowbreak -0.073~\pm~0.026)$~mas~yr$^{-1}$ from 66 galaxies for HST over B5. In the E3 field, the F360M $\mu_\alpha^*$ is larger than the HST-to-HST $\mu_\alpha^*$ by 0.18~mas~yr$^{-1}$ on A5 ($+0.247~\pm~0.101$ against $+0.067~\pm~0.030$~mas~yr$^{-1}$; 13 and 32 galaxies) and by 0.14~mas~yr$^{-1}$ on B5 ($+0.201~\pm~0.142$ against $+0.064~\pm~0.041$~mas~yr$^{-1}$; 8 and 27 galaxies).

The LW measurements use fewer galaxies than the same footprints in HST, because many galaxies returned large PMs in the LW filters alone: in F430M, the cut on PMs larger than 1~mas~yr$^{-1}$ (Section~\ref{ssec:combine}) removes 85 objects, against one in the HST-to-HST measurement over the same footprint. These per-galaxy offsets are correlated between F250M and F430M (correlation coefficients of 0.57 and 0.69 in $\mu_\alpha^*$ and $\mu_\delta$ for 84 galaxies measured in both), but only weakly between F430M and HST (0.10 and 0.04), and so are not noise.
We suspect that this is in large part an effect of the intrinsic changes in the galaxies between F814W and the LW filters (Figure~\ref{fig:galsbywavelength}), which shift the centers fit with the F814W templates. Our injection tests cannot measure such changes, because an injected galaxy has the same shape in every filter. However, in the SW filters they show that offsets of this kind, which are common to two NIRCam filters and absent from the HST-to-HST measurement, also arise from the measurement itself (Section~\ref{ssec:precision}), and so the correlation between F250M and F430M does not by itself show that the galaxies have changed.

So why are the LW images still able to recover the PM, then? Even though there are fewer usable galaxies per SW detector area, the big advantage that the LW filters have is that their single chip covers an area larger than all four of the SW chips under it (since there are no chip gaps). Importantly, since we processed each detector individually, the LW detectors had more stars than any of the SW detectors and so the transformations are less prone to local systematics in the field. Though the large uncertainties may hurt the scientific utility of these results at the distance of WLM, there seems to be promise for these to be incorporated in measurements for targets that are either closer or for which the time baseline is longer. Alternatively, finding a way to combine these data with the SW data may allow for reducing the SW systematics while retaining their higher galaxy counts.

\subsection{Comparative Precision of E2 and E3} \label{ssec:precision}

The B2 detector, which overlaps the ACS field fully and holds the most galaxies, allows the most direct comparison of the precision of the two second epochs, E2 (in F090W and F150W) and E3 (in F200W). Since the two measurements have different baselines, we compare the PM uncertainty multiplied by the baseline, which is the equivalent uncertainty in position.

Before the $\chi^2$ scaling, the F090W measurement (30 galaxies) has an equivalent position uncertainty of 0.17 and 0.18~mas, about half the 0.37~mas of the HST-to-HST measurement over the same area (32 galaxies), and the F150W measurement reaches 0.23~mas in each component with 23 galaxies. The same holds on B1 (0.34 and 0.43~mas for F090W, against 0.78~mas for HST).
The NIRCam second epoch therefore measures the positions of these galaxies more precisely, though with about four times the total exposure time of the second ACS epoch (Table~\ref{tab:obs}).
However, the HST-to-JWST measurements have $\chi^2_\nu = 3.0$ and 2.0 in $\mu_\alpha^*$ and $\mu_\delta$ (F090W on B2), against 0.70 and 0.78 for HST, and after scaling by $\sqrt{\chi^2_\nu}$ the F090W equivalent uncertainties (0.30 and 0.26~mas) are no larger than the 0.37~mas of HST.
At equal baselines and over the area of one SW detector, then, an F090W second epoch gives at least the same PM precision as a second ACS epoch, with the excess per-galaxy scatter shown in Section~\ref{ssec:res_overview}, rather than the position precision, as the limit. 

The F200W measurements of E3 are only comparable to the HST-to-HST measurements over the same areas before the scaling (0.36 and 0.39~mas on A3 from 11 galaxies, against 0.33 and 0.42~mas from 19), and have larger $\chi^2_\nu$ (4.5 and 4.5 on A3).

Three things other than the filter differ between these measurements, and should be kept apart when they are compared: the baseline, the exposure time, and the number of galaxies. The E2 baseline (7.02~years) is shorter than those of E3 (8.93~years for ACS and 9.02~years for NIRCam), and so the same uncertainty in position is a PM uncertainty that is 1.3 times larger for E2. The exposure times differ by more, but matter less. The E3 F200W exposures are a fifth as long in total as the E2 F090W exposures (Table~\ref{tab:obs}), yet the median uncertainty of a single galaxy, expressed in position, is nearly the same in the three SW filters (about 1.3~mas, against 2.6--2.8~mas for HST-to-HST), and few NIRCam galaxies are removed by our cut on the PM uncertainty (3--5 per filter, against 23--29 for the second ACS epoch).

\begin{deluxetable*}{lccccc}
\tablewidth{0pt}
\tabletypesize{\footnotesize}
\tablecaption{Second Epochs over the SW Detector Areas: Exposure Time, Baseline, and Galaxy Selection\label{tab:epochs}}
\tablehead{
\colhead{} & \multicolumn{3}{c}{E2 B1--B3} & \multicolumn{2}{c}{E3 A3, A4, B3, B4} \\
\cline{2-4} \cline{5-6}
\colhead{} & \colhead{NIRCam F090W} & \colhead{NIRCam F150W} & \colhead{ACS F814W} & \colhead{NIRCam F200W} & \colhead{ACS F814W}
}
\startdata
Total exposure time (s) & 30,148 & 23,448 & 7278 & 5840 & 7278 \\
Baseline (yr) & 7.02 & 7.02 & 8.93 & 9.02 & 8.93 \\
Galaxies with a measured PM & 120 & 126 & 106 & 96 & 86 \\
Removed: PM uncertainty $\geq 1$~mas~yr$^{-1}$ & 3 & 5 & 29 & 4 & 23 \\
Removed: $|\mu| > 1$~mas~yr$^{-1}$ at $>3\sigma$ & 19 & 55 & 1 & 44 & 1 \\
Removed: quality of fit & 43 & 27 & 19 & 9 & 6 \\
Removed: $f_\star \geq 0.5$ & 8 & 7 & 10 & 6 & 9 \\
Removed: 3$\sigma$ clip & 4 & 2 & 0 & 9 & 0 \\
Galaxies used & 43 & 30 & 47 & 24 & 47 \\
Median uncertainty per galaxy (mas) & 1.25 & 1.31 & 2.81 & 1.28 & 2.58 \\
\enddata
\tablecomments{The ACS columns are the HST-to-HST measurements, whose second epoch is E3, over the same detector areas. Counts are summed over the detectors, and the cuts are those of Section~\ref{ssec:combine}, in the order in which they are applied. The uncertainty per galaxy is the median PM uncertainty per component of the galaxies used, multiplied by the baseline.}
\end{deluxetable*}

The number of galaxies is the larger difference (Table~\ref{tab:epochs}). Over E2 B1--B3, 43 (F090W) and 30 (F150W) galaxies pass our selection, against 47 for HST-to-HST over the same area, and over the four E3 detectors only 24 pass in F200W, again against 47. The NIRCam measurements lose their galaxies to the cut on PMs larger than 1~mas~yr$^{-1}$ and to the cut on the quality of the fits (Section~\ref{ssec:combine}), which together remove 52\% (F090W), 65\% (F150W), and 55\% (F200W) of the galaxies that have a measured PM, against 8--19\% for HST-to-HST. The first of these cuts alone removes 16\% of them in F090W, 44\% in F150W, and 46\% in F200W, against 1\% for HST-to-HST. A background galaxy cannot have a PM of this size, which is nearly ten times the reflex of WLM's motion, and so every galaxy that this cut removes has a measured PM that is not physical. Their median PMs are 1.5--2.5~mas~yr$^{-1}$, which means that their NIRCam positions are offset from where their F814W templates place them by a median of 16~mas, or half of an SW pixel. The fraction of such galaxies follows the filter and not the exposure time, as F150W, with four times the exposure time of F200W, loses as many.

Our injection tests separate two contributions to these offsets. Because our injected galaxies have the same shape in every filter, any offset that these tests show comes from a property of the measurement (either in terms of methodology or the instrument). In the injected realizations with a motion close to that of WLM (Section~\ref{ssec:injection}), the same cut removes 11\% of the injected galaxies in F090W, 16\% in F150W, and 13\% in F200W, compared to 2\% for HST-to-HST (regardless of whether the galaxies are placed away from stars). The measurement itself can therefore account for more than half of the galaxies that F090W loses to this cut, but only for about a third of those that F150W and F200W lose. Since the number of galaxies cut grows with the separation in wavelength from F814W, our top suspicion is that the remainder is related to intrinsic changes in galaxy morphology between F814W and the near-infrared (NIR) filters.

However, for the galaxies that are not cut, evidence seems to suggest that the extra variance between the HST-to-HST and HST-to-JWST measurements is mostly not related to color-based changes in galaxy profiles. For the galaxies that are used in the F090W, F150W, and HST-to-HST measurements of the same area, the covariances between the three sets of positions isolate a per-galaxy offset that is common to the two NIRCam filters and absent from the HST-to-HST measurement. For the injected galaxies, this offset has a variance of 2.1--2.7~mas$^2$ (a scatter of about 1.5~mas)---with or without star avoidance---and the 26 real galaxies give $2.2 \pm 1.3$~mas$^2$. An offset of this size accounts for most of the excess $\chi^2_\nu$ of the HST-to-JWST measurements. Because the injected galaxies show it, it must be produced by the measurement and not by the galaxies. We have not identified its origin. The real galaxies show a further $0.7 \pm 0.4$~mas$^2$ that differs between F090W and F150W, which the injected galaxies do not show. Whatever their causes, the cut can only remove the galaxies in which the offset is large, and the same effect may be present at a lower level in the galaxies that remain. For those galaxies, the mean offsets from the HST-to-HST PMs of the same galaxies are about the same in the three filters (Section~\ref{ssec:res_overview}). We therefore find no evidence that the F200W PMs are more biased than those of F090W. They are less precise because fewer galaxies survive at wavelengths farther from F814W, and they would be less precise still without their longer baseline.

\section{Discussion and Outlook} \label{sec:discussion}

\subsection{What the Cross-observatory Comparison Shows} \label{ssec:disc_comparison}

In this paper, we have described the measurement of absolute systemic PMs; i.e., PMs measured by defining an external, stationary reference frame (here, specifically, being background galaxies). Knowing the absolute motions of galaxies, and being able to reconstruct their orbits relative to other galaxies, is essential for making measurements and testing predictions in cosmological contexts, at the scale of galaxy systems and making connections to large-scale structure.
Relative PMs between HST and JWST---that is, motions of stars relative to other stars in the same field---can be measured much more precisely (e.g., \citealt{Libralato2023, Ziliotto2023}), in part because relative measurements rely only on measuring positions of stars using well-understood stellar PSFs.

For WLM, we were able to test each step of the absolute measurement. Our HST-to-HST PM agrees with the three Gaia-based PMs to within 0.4$\sigma$ (Section~\ref{ssec:systemic}). The combined HST-to-JWST PMs in F090W and F150W agree with the HST-to-HST PM of the same area to within 0.02 and 0.07~mas~yr$^{-1}$ ($\leq$0.9$\sigma$) and with our adopted full-field ACS PM to within 0.04~mas~yr$^{-1}$ (Section~\ref{ssec:res_overview}). We therefore find that a NIRCam second epoch can recover the absolute PM of a galaxy at $\sim$1~Mpc from an archival ACS/WFC first epoch with a sufficient time baseline.

The agreement is best where the NIRCam filter is closest in wavelength to the ACS/WFC filter.
The primary difference between HST-to-JWST results, however, seems to be closely correlated with the number of background galaxies that can be used for a given measurement.
In the measurement using F090W, we lose $\sim16$\% of the field's background galaxies (that have recovered positions) to our cut on PMs larger than 1~mas~yr$^{-1}$ (Section~\ref{ssec:combine}), which removes background galaxies whose measured PMs would imply physically implausible speeds at WLM's distance. This is compared to 44\% being cut in F150W; 46\% in F200W; 1\% for HST-to-HST; and, in the LW filters, most galaxies being lost through this cut (Section~\ref{ssec:res_lw}).
Our injected galaxies, which have the same shape in every filter, reproduce part of these losses ($\sim11$--16\%), and most of the excess scatter of the galaxies that we keep (Section~\ref{ssec:precision}).
This implies, therefore, that at least part of this loss comes from systematics related to the measurement, not from the galaxies' properties, and we expect that this portion of the systematic can be reduced as the astrometric calibration between ACS/WFC and NIRCam is better understood.
The remainder, which seems to grow with wavelength, is likely owed to the fact that the appearance of a galaxy is a function of wavelength, so a template built in F814W places the center of the same galaxy differently than in a redder filter.

These comparisons can also only test the HST-to-JWST PMs as far as their uncertainties allow.
The difference between the combined F090W PM and the HST-to-HST PM of the same area has an uncertainty of about 0.06~mas~yr$^{-1}$ in each component, and the PMs of the galaxies common to both measurements differ by 0.05--0.17~mas~yr$^{-1}$ (0.8--2.2$\sigma$), with the same sign in all three SW filters (Section~\ref{ssec:res_overview}).
A cross-observatory systematic at the level of $\sim$0.05~mas~yr$^{-1}$ therefore cannot be totally ruled out with these data.
However, it is also relevant that the baselines for these measurements are shorter than what we would typically suggest for absolute PM measurements at 1~Mpc.
Over the 7.02~years between the ACS/WFC and NIRCam epochs, the uncertainty on our combined E2 F090W PM corresponds to a total displacement of about 0.3~mas, or less than 0.01~ACS/WFC pixels.
Whether a systematic offset between the two observatories comes from the templates or from the calibration of the two frames, it is an offset in position that does not grow with time, and so its effect on the PM decreases in proportion to the temporal baseline. In fact, the per-galaxy position uncertainties calculated from routine error propagation are lower with the NIRCam second epochs than with the HST second epoch (a median of 2.6--2.8~mas per galaxy for HST-to-HST, versus about 1.3~mas for F090W and F150W; Table~\ref{tab:epochs}). Prior to scaling the uncertainties by $\sqrt{\chi^2_\nu}$ (Section~\ref{ssec:combine}), the propagated PM uncertainty of the E2 F090W measurement is smaller than that of the HST-to-HST measurement of the same sky area (0.022--0.024 versus 0.030--0.032~mas~yr$^{-1}$), despite its shorter temporal baseline and slightly lower number of galaxies (43 versus 47). However, for HST-to-HST in the combined SW E2 area, we calculate $\chi^2_\nu=0.8$, versus $\chi^2_\nu = 2.6$ for F090W. It is this scaling, together with the stationary and moving bias corrections from our injection tests, that leaves the F090W measurement as the less precise of the two.
Longer baselines therefore reduce both the random and the systematic uncertainties, and so there is reason to believe that, with longer baselines, equal-area cross-observatory PM measurements should converge to the precision of the HST-to-HST measurement. Still, how fast that convergence happens relies on further work to understand these systematics, but there is no reason to believe that the size of the systematic implied by our data is an intrinsic feature of cross-observatory astrometry. The PMs of future Gaia data releases will provide a test of these offsets that is independent of HST.

\subsection{Improving HST-to-JWST PMs} \label{ssec:disc_improve}

For programs that plan a NIRCam second epoch to archival HST imaging, our results suggest the following.
\begin{enumerate}
    \item Observe in the NIRCam filter that is closest in wavelength to the first-epoch filter. For an F814W first epoch, this is F090W.
    \item Cover as much of the first-epoch field as possible with the SW detectors. Our uncertainties are set by the number of galaxies that pass our selection, and a single NIRCam pointing leaves gaps between its detectors and modules. Because we fit each detector separately, the most favorable targets are those compact enough to place most of their stars on a single SW detector. It may even be worth requesting orientation restrictions to ensure maximal coverage of background galaxies pre-identified in archival images.
    \item For the NIRCam epoch, prioritize a higher number of dithered exposures over deeper individual exposures (e.g., eight 500~s exposures versus four 1000~s exposures). For HST-to-HST PMs, long exposure times are preferred to ensure adequate SNR for background galaxies in each individual exposure. However, the position uncertainty of a galaxy in a single NIRCam exposure is nearly the same in our three SW filters (1.1--1.4~mas), despite their exposure times differing by a factor of five. Few NIRCam galaxies are lost solely to our cut on the PM uncertainty ($\lesssim 5$ per filter).
    \item In the immediate term, if the most precise PM measurement is the primary science case, it is preferable to use HST for the second epoch while ACS/WFC remains available. The first-epoch filter, pointing, and orientation can be repeated, allowing the use of the whole field and for systematics to be minimized.
\end{enumerate}

There is also room to improve the method itself.
As a few examples:
We built our background galaxy catalog by eye because the automated selection by color that we tested performed poorly, at least in these crowded fields (Section~\ref{ssec:catalogs}). A selection that works in such fields would enlarge the sample and make it reproducible. The offsets of galaxies between F814W and the NIRCam filters could be calibrated against their colors and morphologies, using fields that both observatories have imaged deeply. 
Subtracting the neighboring stars recovered only part of the PM that is lost to crowding in our tests (Section~\ref{ssec:crowding}), and a better model of the crowded scene could remove the need for our scale correction (Section~\ref{ssec:injection}).
Stars with Gaia PMs and QSOs, where a field has them, give reference frames that are independent of the galaxies, and can be used to check them (as in \citealt{Warfield2026}).
Many of our solutions to problems we encountered---for instance, blurring the NIRCam images with a Gaussian kernel to bring them to/below ACS/WFC resolution---were adequate for making the measurement possible, but also involved throwing away potentially useful data from NIRCam's higher resolution.
Though, traditionally, HST-to-HST measurements have stuck to using one filter in each epoch (usually the same filter), it may be worth exploring how the information from multiple filters can contribute to position measurements in a given epoch, especially considering that each SW observation is paired with an LW observation.
Finally, a sub-pixel accurate astrometric calibration that ties the NIRCam detectors to each other may allow all of the detectors to be fit together, as are the two ACS/WFC chips.

\subsection{The Future of PMs in the Local Volume} \label{ssec:disc_future}

One seemingly unavoidable obstacle (but exciting challenge) on the horizon is our ability to follow up archival observations with second-epoch imaging where the existing temporal baselines are too short.
The most immediate case is that of the satellites of M31, for most of which deep first-epoch imaging was taken between 2019 and 2021 \citep{Savino2022, Savino2025}.
The results that we have presented in this paper can be used to give us some idea about what we might expect in terms of baselines needed and likely (or possible) precision.
Our second ACS/WFC epoch has six exposures against the twenty-six of the first, and carries about 70\% of the variance of our HST-to-HST PM.
If a second epoch was as deep and well-sampled as our first WLM epoch, also using HST, the positional precision of that measurement would improve from 0.16 to 0.11~mas.
With two such epochs, a tangential velocity uncertainty of 60~km~s$^{-1}$ (per component) would require a baseline of about six years at the distance of M31 (about ten years with data identical to our WLM HST epochs).
In the last column of Table~\ref{tab:lv_census}, we tabulate this baseline estimate for the M31 satellites, as well as for a number of other groups of galaxies within 5~Mpc (from the Local Volume Database; \citealt{Pace2025}). 

Until the systematics that we find between HST and JWST are understood more fully (Section~\ref{ssec:disc_comparison}), these baselines should be taken as the optimistic case for cross-observatory measurements. Programs that depend on cross-observatory measurements are already underway. JWST-GTO 4514 \citep{vanderMarel2023}, for example, will measure the PMs of the M31 satellites Andromeda~XIV, XVII, and XXII against either archival HST imaging or earlier JWST imaging, and the archival HST epochs that already exist for many LG systems form a legacy sample for follow-up with JWST and future facilities \citep{Sohn2026}.

\begin{deluxetable*}{lccccccc}
\tablecaption{Galaxies within 5~Mpc: First-epoch Imaging and Required Temporal Baselines \label{tab:lv_census}}
\tablehead{\colhead{Group} & \colhead{$N_{\rm gal}$} & \colhead{$D$} & \colhead{$N_{\rm first}$} & \colhead{$N_{\rm survey}$} & \colhead{$N_{\rm none}$} & \colhead{JWST Time} & \colhead{$\Delta t_{60}$} \\
\colhead{} & \colhead{} & \colhead{(Mpc)} & \colhead{} & \colhead{} & \colhead{} & \colhead{(hr)} & \colhead{(yr)}}
\startdata
MW satellites & 20 & 0.1 & 11 & 9 & 6 & 22 & 1.0 \\
M31 satellites & 43 & 0.8 & 38 & 0 & 5 & 18 & 6.4 \\
LG field & 14 & 0.7 & 13 & 4 & 1 & 4 & 6.4 \\
NGC 3109 association & 6 & 1.4 & 6 & 2 & 0 & 0 & 12 \\
Sculptor filament & 35 & 3.5 & 25 & 34 & 1 & 4 & 30 \\
IC 342 / Maffei & 9 & 3.2 & 4 & 0 & 5 & 18 & 27 \\
M81 group & 59 & 3.7 & 38 & 42 & 5 & 18 & 31 \\
Cen A / M83 & 118 & 3.7 & 31 & 0 & 87 & 313 & 31 \\
CVn I / M94 & 50 & 4.3 & 25 & 0 & 25 & 90 & 37 \\
Other field & 63 & 3.5 & 40 & 12 & 17 & 61 & 29 \\
\hline
Total & 417 & 3.6 & 231 & 103 & 152 & 547 & 31 \\
\enddata
\tablecomments{Objects in the Local Volume Database \citep{Pace2025} within 5~Mpc, including 88 unconfirmed candidates, and excluding the 48 MW satellites whose Gaia-based PMs already have uncertainties below 60~km~s$^{-1}$ per component and the massive hosts of each group. $D$: median distance. $N_{\rm first}$: galaxies with ACS/WFC, Wide Field Camera 3, NIRCam, or NIRISS imaging that reaches a signal-to-noise ratio of 10 one magnitude below the tip of the red giant branch. $N_{\rm survey}$: galaxies inside a Roman core survey or the Euclid DR1 area. $N_{\rm none}$: galaxies with neither (69 of the 152 are candidates). JWST time: the charged time for one NIRCam pointing per galaxy in $N_{\rm none}$, with eight dithers of 408~s in each of F090W and F150W (3.6~hr per galaxy). $\Delta t_{60}$: the temporal baseline at which two ACS/WFC epochs of 26 exposures each reach 60~km~s$^{-1}$ per component at the median distance of the group, from the positional precision of our HST-to-HST PM (0.16~mas with 26 and 6 exposures) scaled to that depth (0.11~mas). This is a single-instrument case, and optimistic for cross-observatory measurements.}
\end{deluxetable*}

This is even more relevant for resolved populations beyond the LG. Of the 417 galaxies in Table~\ref{tab:lv_census} that do not already have a precise Gaia-based PM, 231 have HST or JWST imaging that is deep enough to serve as a first epoch. In contrast, 152 galaxies have none and lie outside the Roman core surveys and the first Euclid data release.
A single NIRCam pointing for each of these 152 galaxies would take about 550~hr of JWST time. From our extrapolation of the WLM HST-to-HST measurement, the groups at 3--4~Mpc need baselines of about 30~years to reach 60~km~s$^{-1}$.\footnote{Baselines of twenty or more years have been estimated for the Centaurus~A and M81 systems \citep[e.g.,][]{Bennet2025CenA}. For any particular galaxy, the true baseline needed will be dependent on the design of the first-epoch observations, the properties of the targeted galaxy (including number of member stars and crowding), and the number of usable background galaxies or other reference frame sources that happen to be available in those images.}
These baselines shorten in proportion to any gain in the precision of the two epochs, for example from more exposures or more background galaxies (Section~\ref{ssec:disc_improve}), but they are in every case long enough that the first epochs must be taken soon.

Perhaps the biggest remaining question in terms of observing strategy, then, is which telescopes should be used for new first-epoch observations. JWST, owing to an excellently executed launch, has the potential to be operational into the 2040s or beyond, when its propellant (needed to remain in orbit around L2) will be depleted.\footnote{NASA has estimated that Webb carries enough propellant for ``significantly more than a 10-year science lifetime'' (NASA Webb Mission Team, ``NASA Says Webb's Excess Fuel Likely to Extend its Lifetime Expectations,'' 2021 December 29; \url{https://science.nasa.gov/blogs/webb/2021/12/29/nasa-says-webbs-excess-fuel-likely-to-extend-its-lifetime-expectations/}).}
It may therefore have enough time to be used for PM measurements of resolved populations at distances of up to 5~Mpc, but only if the first-epoch observations are executed now. The Nancy Grace Roman Space Telescope (launched in 2026 August) also observes in the NIR, and so has strong potential to work in tandem with JWST for either first or second epochs. However, as Roman is a survey telescope, this depends heavily on the programs selected for its General Astrophysics Surveys.
Euclid will image many of these systems as part of its wide survey \citep{Euclid2025}---though with pixels of 0\farcs1, twice the size of those of ACS/WFC, and a correspondingly broader PSF. The Chinese Space Station Survey Telescope \citep{CSST2026}, scheduled for launch in 2027, will survey 17,500~deg$^2$ in the optical from space, and the proposed GaiaNIR mission \citep{Hobbs2021} would, in combination with Gaia, give much improved PMs for the stars common to both. Further in the future, NASA's next flagship mission, the Habitable Worlds Observatory (HWO), is planned for the 2040s. HWO will mark a return to UV--optical imaging, which may make the case to still prefer HST for new first-epoch observations of distant galaxies.

\section{Summary} \label{sec:conclusions}

In this paper, we have presented absolute PM measurements of the isolated LG galaxy WLM from HST/ACS imaging in 2015 and 2024 and from JWST/NIRCam imaging in 2022 and 2024, using background galaxies as the reference frame, including the first absolute systemic PMs of a galaxy measured by combining HST and JWST imaging. We summarize our results as follows.
\begin{enumerate}
    \item From the two ACS/WFC epochs, we measure a systemic PM of $(\mu_\alpha^*, \mu_\delta) = (+0.087 \pm 0.019,\allowbreak -0.079 \pm 0.017)$~mas~yr$^{-1}$ from 145 background galaxies over 8.93~years, or $(V_\alpha, V_\delta) = (+409 \pm 88,\allowbreak -372 \pm 79)$~km~s$^{-1}$ at 989.5~kpc, and $(+287 \pm 88,\allowbreak -167 \pm 79)$~km~s$^{-1}$ after correcting for the solar motion. This agrees to within 0.4$\sigma$ with the three previous Gaia-based PMs, and is the most precise PM of WLM and the only one that does not rely on Gaia (Section~\ref{ssec:systemic}).
    \item To measure PMs between ACS/WFC and NIRCam, we modified the template-fitting method of \citet{Warfield2026}, which builds on that of \citet{Mahmud2008} and \citet{Sohn2012}. We fit each NIRCam detector separately, convolve the F814W templates with kernels that account for the difference in PSF, remove background galaxies whose stamps are dominated by the light of neighboring stars, and correct the PMs with injection tests for an additive bias and for a scale factor of 0.77--0.96 (Section~\ref{sec:methods}).
    \item From the 2022 NIRCam epoch, 7.02~years after the first ACS/WFC epoch, we measure PMs of $(+0.081 \pm 0.045,\allowbreak -0.113 \pm 0.046)$~mas~yr$^{-1}$ in F090W (43 galaxies) and $(+0.084 \pm 0.062,\allowbreak -0.051 \pm 0.057)$~mas~yr$^{-1}$ in F150W (30 galaxies). Both agree with the HST-to-HST PM of the same area and with our adopted PM to within 1$\sigma$. The F200W PM from the 2024 NIRCam epoch differs from the HST-to-HST PM of its area by 1.8 and 2.1$\sigma$ (Section~\ref{ssec:res_overview}).
    \item The NIRCam measurements lose 16\% (F090W), 44\% (F150W), and 46\% (F200W) of their background galaxies to PMs that are not physical, against 1\% for HST-to-HST, and the HST-to-JWST PMs of the galaxies that remain differ from their HST-to-HST PMs by 0.05--0.17~mas~yr$^{-1}$ (0.8--2.2$\sigma$). Cross-observatory systematics of $\sim$0.05~mas~yr$^{-1}$ therefore cannot be ruled out over a baseline of 7~years. Background galaxies injected with the same shape in every filter reproduce a part of these losses and most of the excess scatter, which therefore come in part from the measurement and not from the galaxies' properties (Sections~\ref{ssec:precision} and \ref{ssec:disc_comparison}).
    \item In the LW filters, most background galaxies are lost to the same cut. However, the F250M and F430M PMs are consistent with the HST-to-HST PM of the same footprint, with uncertainties of about 0.09~mas~yr$^{-1}$ (Section~\ref{ssec:res_lw}).
    \item At equal baselines and over the area of one SW detector, an F090W second epoch gives at least the PM precision of a second ACS/WFC epoch. For a NIRCam second epoch, we suggest the filter closest in wavelength to that of the first epoch, and as much coverage of the first-epoch field as possible (Section~\ref{ssec:disc_improve}).
\end{enumerate}

Putting this study in a larger context, the methods in this paper are not limited to ACS/WFC and NIRCam and are applicable to any cross-observatory, template-based astrometric measurement. The need for longer baselines means that data from a variety of telescopes will be increasingly essential, with the pay-off of some observations coming past the careers of many astronomers currently in the field. If the astronomical community sees measurements of PMs and galactic dynamics as important and worthwhile, then they must be considered seriously in the long-term planning of current and future missions. Fortunately, we are at a transition point where the observatories that defined the previous era and those that will define the next are operating at the same time. The measurements in this paper are one piece in bridging these observatories together, and it is therefore the right moment to determine the role that we would like high-precision PM measurements to have.

\begin{acknowledgments}
JTW acknowledges support from the Jefferson Scholars Foundation Graduate Fellowship.

This project was carried out in collaboration with the High-resolution Space Telescope PROper MOtion (HSTPROMO) Collaboration (\url{https://www.stsci.edu/~marel/hstpromo.html}), a set of projects aimed at improving our dynamical understanding of stars, clusters, and galaxies in the nearby Universe through measurement and interpretation of PMs from HST, JWST, and Gaia. We thank the collaboration members for the sharing of their ideas and software.

This research is based on observations made with the NASA/ESA Hubble Space Telescope obtained from the Space Telescope Science Institute, which is operated by the Association of Universities for Research in Astronomy, Inc., under NASA contract NAS 5-26555. These observations are associated with programs GO-13768 (PI: D. Weisz) and GO-17473 (PI: J. Warfield). Support for programs HST-GO-17473 (PI: J. Warfield) and HST-AR-17026 (PI: D. Weisz) was provided by NASA through grants from the Space Telescope Science Institute, which is operated by the Association of Universities for Research in Astronomy, Inc., under NASA contract NAS 5-26555.
This work is also based on observations made with the NASA/ESA/CSA James Webb Space Telescope. The data were obtained from the Mikulski Archive for Space Telescopes at the Space Telescope Science Institute, which is operated by the Association of Universities for Research in Astronomy, Inc., under NASA contract NAS 5-03127 for JWST. These observations are associated with programs ERS-1334 (PI: D. Weisz) and GO-4570 (PI: J. Warfield). Support for program JWST-GO-4570 (PI: J. Warfield) was provided by NASA through a grant from the Space Telescope Science Institute, which is operated by the Association of Universities for Research in Astronomy, Inc., under NASA contract NAS 5-03127.
\end{acknowledgments}

\begin{contribution}

JTW was responsible for the data analysis and measurement, and writing and submitting the manuscript. STS and JA developed the software for HST-to-HST measurements---including for setting the reference frame and fitting galaxy templates---on which the present work builds further.
All other authors read and provided feedback on the manuscript.
This paper was constructed and written by the authors, but portions were edited with the assistance of Claude Opus v5.4 and 5.5, large language models developed by Anthropic, to improve readability, to conform the draft to the AAS style guide, as well as for routine technical tasks such as code and plot formatting. All content was reviewed and verified by the authors.

\end{contribution}

\facilities{HST(ACS), JWST(NIRCam)}

\software{astropy \citep{astropy2013, astropy2018, astropy}; astroquery \citep{Ginsburg2019}; DOLPHOT \citep{Dolphin2000, Dolphin2016, Weisz2024}; HubPUG \citep{Warfield2023}; hst1pass \citep{Anderson2022}; jwst1pass \citep{Libralato2023, jwst1pass_readme}; the JWST Science Calibration Pipeline \citep{Bushouse2025}; Matplotlib \citep{Hunter2007}; NumPy \citep{Harris2020}; pandas \citep{Mckinney2010}; SAOImage DS9 \citep{SAO2000}; SciPy \citep{Virtanen2020}; xym2mat, xym2bar, and img2stk (J. Anderson); hips2fits,\footnote{\url{https://alasky.cds.unistra.fr/hips-image-services/hips2fits}} a service provided by CDS}

\bibliography{bib}{}
\bibliographystyle{aasjournalv7.1}

\end{document}